\documentclass[aps,prx,reprint,superscriptaddress,nofootinbib,longbibliography,floatfix]{revtex4-2}

\usepackage[T1]{fontenc}
\usepackage[utf8]{inputenc}
\usepackage{amsmath,amssymb,mathtools}
\usepackage{bm}
\usepackage{graphicx}
\usepackage{xcolor}
\usepackage{hyperref}

\hypersetup{colorlinks=true,linkcolor=blue,urlcolor=blue,citecolor=blue}

\newcommand{\ii}{\mathrm{i}}
\newcommand{\Tr}{\operatorname{Tr}}

\begin{document}

\title{Liouvillian Quasiparticle Spectroscopy}

\author{Kai Li}
\email[Contact author: ]{kaili24@foxmail.com}
\affiliation{RIKEN Center for Emergent Matter Science (CEMS),
Wako, Saitama 351-0198, Japan}
\author{Masahito Ueda}
\email[Contact author: ]{ueda@cat.phys.s.u-tokyo.ac.jp}
\affiliation{Department of Physics, The University of Tokyo,
7-3-1 Hongo, Bunkyo-ku, Tokyo 113-0033, Japan}
\affiliation{Institute for Physics of Intelligence, The University of Tokyo,
7-3-1 Hongo, Bunkyo-ku, Tokyo 113-0033, Japan}
\affiliation{Fundamental Quantum Science Program (FQSP), TRIP Headquarters, and RIKEN Center for Emergent Matter Science (CEMS), RIKEN, Wako 351-0198, Japan}

\begin{abstract}
Quasiparticles play a central role in understanding
closed quantum many-body systems, as they reduce complex many-body problems
to effective single-particle descriptions and connect directly to spectroscopic measurements.
In open quantum systems, the interplay between coherent dynamics and dissipation
gives rise to unique phenomena, such as the Liouvillian
skin effect and strong-to-weak spontaneous symmetry breaking.
This raises a fundamental question: how can the quasiparticle picture 
be extended to open quantum systems to account for such phenomena and 
enable the experimental probing of quasiparticle spectra?
Here we introduce coherent and incoherent Liouvillian
quasiparticles, defined by responses to perturbations
acting on one side and both sides of the density matrix, respectively.
We propose spectroscopy protocols that couple the system
to auxiliary levels---either coherently or via a dissipative pulse---to
measure the corresponding complex quasiparticle energies.
Numerical simulations of Liouvillians with skin effects 
validate these protocols, revealing nontrivial spectral winding 
associated with unidirectional transport in real space.
Notably, coherent and incoherent quasiparticles can propagate
in opposite directions within the same system.
Furthermore, in models exhibiting weak and strong-to-weak $U(1)$ spontaneous
symmetry breaking, our results unveil gapless Nambu--Goldstone
modes in the coherent and incoherent quasiparticle spectra, respectively;
the other quasiparticle spectrum remains gapped in each case.
Our work provides a unified framework for theoretically describing
and experimentally probing coherent and incoherent
quasiparticles in open quantum many-body systems.

\end{abstract}

\maketitle

\section{Introduction}

The quasiparticle picture serves as a foundational framework for understanding
quantum many-body systems~\cite{Laughlin2000PNAS,Keimer2017NatPhys}.
Its importance is twofold.
First, it maps complex many-body problems
onto effective single-particle descriptions, thereby simplifying theoretical treatments.
As a prototypical example, Fermi-liquid theory describes the low-energy excitations
of interacting electrons as quasiparticles that carry 
the electron charge and spin but possess renormalized masses
and finite lifetimes~\cite{Landau1957JETP}.
Second, the quasiparticle picture links directly to experimental
observables through spectroscopic methods. These include
angle-resolved photoemission spectroscopy~\cite{Damascelli2003RMP, Sobota2021RMP}
and inelastic neutron scattering~\cite{Dai2015RMP} in solid-state materials,
as well as radio-frequency~\cite{Gupta2003Science,Chin2004Science} 
and Bragg spectroscopy~\cite{Stenger1999PRL, StamperKurn1999PRL}
in ultracold atomic systems.
These techniques offer direct access to quasiparticle spectra,
unveiling signatures of density-wave order~\cite{Tranquada2004Nature},
superfluidity and superconductivity~\cite{Shen1993PRL, Chin2004Science, 
Stewart2008Nature,Schirotzek2008PRL},
and topological band structures~\cite{Hsieh2008Nature}.

Open quantum systems have recently attracted considerable
interest~\cite{Diehl2008NatPhys, Verstraete2009NatPhys,
Daley2014AdvPhys, Sieberer2025RMP}.
The interplay between coherent dynamics and dissipation
gives rise to exotic quantum phenomena, such as dissipative phase
transitions~\cite{Kessler2012PRA, Lee2013PRL, Fitzpatrick2017PRX, Minganti2018PRA},
dissipative superfluidity~\cite{Yamamoto2021PRL, Li2025PRL},
mixed-state topology~\cite{Bardyn2018PRX, Ma2023PRX,
Guo2025PRX, Ma2025PRX, Yue2026NatPhys, Su2026NatPhys},
Liouvillian skin effects~\cite{Song2019PRL, Haga2021PRL},
and exceptional points~\cite{Minganti2019PRA}.
Of particular interest are the distinct forms of spontaneous symmetry
breaking (SSB) associated with strong and weak symmetries~\cite{Buca2012NJP},
including weak SSB and strong-to-weak SSB~\cite{Lieu2020PRL, Lee2023PRXQuantum,
Ogunnaike2023PRL, Sala2024PRB, Moudgalya2024PRXQuantum,
Lessa2025PRXQuantum, Huang2025PRB, Gu2025PRB}.
In closed systems, SSB is intimately linked to the emergence of quasiparticles.
For example, the spontaneous breaking of continuous 
global symmetries gives rise to gapless Nambu--Goldstone 
excitations~\cite{Goldstone1961NuovoCimento, Nambu1961PR, Goldstone1962PR}.
Given the unique nature of SSB in open quantum systems, it is natural to ask
how the concept of quasiparticles can be generalized
to open quantum systems and how they can capture the physics
underlying these diverse forms of SSB.

Previous work introduced incoherentons, which are bound states
between the ket and bra degrees of freedom, to describe decoherence
processes in open quantum systems~\cite{Haga2023PRR}.
While this picture provides valuable insights into quasiparticles 
intrinsic to open quantum systems, it is not formulated in terms of response
functions and thus lacks a direct connection to spectroscopy.
Response theories more closely related to spectroscopic measurements have
been developed for open quantum systems via weak perturbations 
to the Liouvillian~\cite{Chetrite2012JStatPhys, CamposVenuti2016PRA,
Sieberer2016RPP, Albert2016PRX, Konopik2019PRR}.
However, how to establish a clear link between these response functions 
and physically relevant quasiparticles remains elusive, 
especially for quasiparticles intrinsic to open quantum systems.
This gap motivates the development of a unified framework that encompasses 
both the theoretical description
and the spectroscopic measurement of quasiparticles in open quantum systems.

In this work, we introduce the concept of coherent and incoherent
Liouvillian quasiparticles and propose spectroscopy protocols
to measure their complex energies.
We define both types of quasiparticle via the poles
of the corresponding response functions.
For coherent quasiparticles, the perturbation in the response function acts on
either the ket or bra side of the density matrix, 
whereas for incoherent quasiparticles, it acts on both sides simultaneously.
These quasiparticles characterize the propagation and decay 
of such perturbations under the Liouvillian dynamics.
In particular, incoherent quasiparticle spectroscopy can reveal density
dynamics that are not captured by the coherent quasiparticle spectrum.
Using linear response theory, we demonstrate that coherent and incoherent
quasiparticle spectra can be measured by coupling the system to auxiliary
levels coherently and via a dissipative pulse, respectively. 
Here, the dissipation is used to probe the open system itself,
in stark contrast to approaches that utilize dissipation to probe
the properties of Hermitian systems~\cite{Pan2020NatPhys,He2026arXiv}.

Numerical simulations of Liouvillians exhibiting the skin effect
demonstrate the feasibility of our protocols,
with the extracted complex spectra showing excellent agreement with those
obtained from direct calculations.
As momentum is varied, the quasiparticle energies wind around some reference points
in the complex plane.
This nontrivial spectral topology is consistent with 
unidirectional transport in real space.
Notably, coherent and incoherent quasiparticles can propagate
in opposite directions within the same system, 
highlighting the distinct properties encoded in their respective spectra.

Furthermore, we employ quasiparticle spectroscopy to investigate
$U(1)$ SSB in open quantum systems.
We introduce two models with weak and strong $U(1)$
symmetries~\cite{Gu2025PRB}.
For the model featuring weak $U(1)$ SSB, our results reveal a gapless
Nambu--Goldstone mode in the coherent quasiparticle spectrum
alongside a gapped incoherent spectrum.
For the model displaying strong-to-weak $U(1)$ SSB, the incoherent spectrum
is gapless, whereas the coherent spectrum remains gapped.
These findings demonstrate that coherent and incoherent quasiparticle spectroscopy 
can distinguish the collective excitations associated with 
different forms of SSB.
In particular, incoherent quasiparticle spectroscopy probes 
excitations intrinsic to open quantum systems.

\section{Coherent Quasiparticles}
\label{sec_coherent_quasiparticles}

We start with the retarded response function
\begin{equation}\label{Eq1}
\chi_k^R (t) = -\ii \Theta (t)
\langle [ \hat{O}_k (t),\hat{O}_k^\dagger (0) ]_\zeta \rangle,
\end{equation}
where $\hat{O}_k (t) = e^{\mathcal{L}_\mathrm{s}^\dagger t}[\hat{O}_k]$ with $\mathcal{L}_\mathrm{s}^\dagger$
being the adjoint of the system Liouvillian $\mathcal{L}_\mathrm{s}$,
$\langle \hat{O} \rangle = \Tr (\hat{O} \rho_\mathrm{s})$, and
$[\hat{O}_1, \hat{O}_2]_\zeta = \hat{O}_1 \hat{O}_2 - \zeta \hat{O}_2 \hat{O}_1$ with $\zeta = +1$ for bosons and $\zeta = -1$ for fermions.
The frequency-domain response function and spectral function are given by
\begin{align}
\chi_k^R (\omega) &= \int_0^\infty d t \, e^{\ii \omega t} \chi_k^R (t),
\\
A_k (\omega) &= -2 \operatorname{Im} \chi_k^R (\omega).
\end{align}
The response function \eqref{Eq1} is a direct generalization of the response function 
in closed systems to open quantum systems.
For $\mathcal{L}_\mathrm{s}[\rho] = -\ii[\hat{H}_\mathrm{s},\rho]$, Eq.~\eqref{Eq1} 
reduces to the response function in closed systems, 
with $\hat{O}_k(t) = e^{\ii\hat{H}_\mathrm{s}t}\hat{O}_k e^{-\ii\hat{H}_\mathrm{s}t}$.

Quasiparticles correspond to poles of the response function in the complex frequency plane.
When the system has well-defined quasiparticles, the response function can be approximated by
\begin{equation}
\chi_k^R (\omega) \approx
\sum_\alpha \frac{a_{k \alpha} + \ii b_{k \alpha}}{\omega - z_{k \alpha}},
\end{equation}
where $\alpha$ labels the quasiparticle branches,
$z_{k \alpha} = \varepsilon_{k \alpha} - \ii \gamma_{k \alpha}$ is the complex quasiparticle energy, 
with $\varepsilon_{k \alpha}$ and $-\gamma_{k \alpha}$ denoting its real and imaginary parts, respectively,
and $a_{k \alpha}$ and $b_{k \alpha}$ are real coefficients.
Consequently, the spectral function can be approximated by
\begin{equation}\label{fiteq}
A_k (\omega) \approx
2 \sum_\alpha
\frac{ a_{k \alpha} \gamma_{k \alpha}
- b_{k \alpha} (\omega - \varepsilon_{k \alpha})
}{
(\omega - \varepsilon_{k \alpha})^2 + \gamma_{k \alpha}^2
}.
\end{equation}
The choice of $\hat{O}_k$ determines which quasiparticle modes are probed by the response function.
Below we take $\hat{O}_k = \hat{c}_{k m}$ to be an annihilation operator for a particle with momentum $k$ in mode $m$.
The corresponding quasiparticles are single-particle-like Liouvillian quasiparticles in the coherent (off-diagonal) sector.

To detect these quasiparticles, we introduce auxiliary levels
and couple them coherently to the system levels~\cite{Li2022PRL}.
The coupling Hamiltonian is
\begin{equation}\label{couplingHamiltonian}
\begin{aligned}
\hat{H}_\mathrm{c} &= \sum_j \left[ \frac{\Omega}{2}
(\hat{a}_j^\dagger \hat{c}_{j m} + \hat{c}_{j m}^\dagger \hat{a}_j) + \omega_\mathrm{a} \hat{a}_j^\dagger \hat{a}_j \right]
\\ &= \sum_k \left[ \frac{\Omega}{2}
(\hat{a}_k^\dagger \hat{c}_{k m} + \hat{c}_{k m}^\dagger \hat{a}_k) + \omega_\mathrm{a} \hat{a}_k^\dagger \hat{a}_k \right],
\end{aligned}
\end{equation}
where $\hat{a}_j$ ($\hat{a}_k$) annihilates a particle on site $j$ (with momentum $k$) in the auxiliary levels, $\Omega$ is the Rabi frequency, 
and $\omega_\mathrm{a}$ is the detuning between the auxiliary and system levels.
The full Liouvillian is given by $\mathcal{L} = \mathcal{L}_\mathrm{s} + \mathcal{L}_\mathrm{c}$ with 
$\mathcal{L}_\mathrm{c}[\rho] = -\ii [\hat{H}_\mathrm{c}, \rho]$.
We then monitor the momentum-resolved auxiliary occupation 
$\hat{N}_{\mathrm{a}, k} = \hat{a}_k^\dagger \hat{a}_k$ and obtain
\begin{equation}
\dot{N}_{\mathrm{a}, k} (t) = \frac{d}{d t} \Tr \! \big[ \hat{N}_{\mathrm{a}, k} \rho (t) \big],
\end{equation}
where $\rho (t) = e^{\mathcal{L} t} [\rho_0]$ is the density matrix evolved by the full Liouvillian.
Here the initial state is given by $\rho_0 = \rho_\mathrm{s} \otimes \rho_\mathrm{a}$.
We assume that $\rho_\mathrm{s}$ is stationary under $\mathcal{L}_\mathrm{s}$ and
$\rho_\mathrm{a}$ is diagonal in the momentum basis.
We note that $\otimes$ denotes the ordinary tensor product for bosons and 
the $\mathbb{Z}_2$-graded tensor product~\cite{Bultinck2017PRB} for fermions.

In general, both the particle transfer from the auxiliary levels to the system levels 
and the reverse process contribute to $\dot{N}_{\mathrm{a}, k} (t)$.
We thus decompose the response as
\begin{equation}
\chi_k^R (t) = \chi_k^+ (t) - \zeta \chi_k^- (t),
\end{equation}
where
\begin{align}
\label{chiplst}
\chi_k^+ (t) &= -\ii \Theta (t) \Tr \! \big\{
\hat{c}_{k m} e^{\mathcal{L}_\mathrm{s} t}
[\hat{c}_{k m}^\dagger \rho_\mathrm{s}]
\big\},
\\
\label{chimnst}
\chi_k^- (t) &= -\ii \Theta (t) \Tr \! \big\{
\hat{c}_{k m} e^{\mathcal{L}_\mathrm{s} t}
[\rho_\mathrm{s} \hat{c}_{k m}^\dagger]
\big\},
\end{align}
with the former corresponding to particle transfer from the 
auxiliary levels to the system levels and the latter to the reverse process.
Similarly, the spectral function can be written as
\begin{align}
A_k (\omega) &= A_k^+ (\omega) - \zeta A_k^- (\omega),
\\
\label{Akpm}
A_k^\pm (\omega) &= -2 \operatorname{Im} \int_0^\infty d t \, e^{\ii \omega t} \chi_k^\pm (t).
\end{align}

Using linear response theory, we obtain
\begin{equation}\label{dotNakt}
\dot{N}_{\mathrm{a}, k} (t) \approx -\frac{\Omega^2}{4} 
\big[ N_{\mathrm{a}, k}^0 A_k^+ (\omega_\mathrm{a}) - (1 + \zeta N_{\mathrm{a}, k}^0) A_k^- (\omega_\mathrm{a}) \big],
\end{equation}
where $N_{\mathrm{a}, k}^0 = \Tr (\hat{N}_{\mathrm{a}, k} \rho_0)$ is the initial auxiliary population at momentum $k$.
The derivation is given in Appendix~\ref{coh_derivation}.
For a fermionic system, the two spectral functions $A_k^{\pm} (\omega_\mathrm{a})$ can be obtained 
by preparing either empty or fully occupied auxiliary levels.
When $N_{\mathrm{a}, k}^0 = 0$, the contribution from $A_k^+ (\omega_\mathrm{a})$ vanishes, and
$\dot{N}_{\mathrm{a}, k} (t) \approx \frac{\Omega^2}{4} A_k^- (\omega_\mathrm{a})$.
When $N_{\mathrm{a}, k}^0 = 1$, the contribution from $A_k^- (\omega_\mathrm{a})$ vanishes, and
$\dot{N}_{\mathrm{a}, k} (t) \approx -\frac{\Omega^2}{4} A_k^+ (\omega_\mathrm{a})$.
For a bosonic system, $A_k^- (\omega_\mathrm{a})$ can similarly be obtained using empty auxiliary levels,
and $A_k^+ (\omega_\mathrm{a})$ can be obtained by using occupied auxiliary levels 
and subtracting the $A_k^-(\omega_\mathrm{a})$ contribution from the signal.

We demonstrate our spectroscopy protocol using the Liouvillian Rice-Mele (LRM) model,
which extends the non-Hermitian Rice-Mele (NHRM) model~\cite{Yi2020PRL} to open quantum systems.
Each unit cell contains two fermionic modes labeled by $\nu = \uparrow, \downarrow$, 
and the system Liouvillian for the LRM model (labeled by a superscript $\mathrm{R}$) is given by 
\begin{equation}
\mathcal{L}_\mathrm{s}^\mathrm{R} [\rho]
= -\ii [\hat{H}_\mathrm{s}^\mathrm{R}, \rho]
+ \sum_{j=1}^L
\Big(
\hat{L}_j^\mathrm{R} \rho (\hat{L}_j^\mathrm{R})^\dagger
- \frac{1}{2} \{ (\hat{L}_j^\mathrm{R})^\dagger \hat{L}_j^\mathrm{R}, \rho \}
\Big),
\end{equation}
where
\begin{equation}
\begin{aligned}
\hat{H}_\mathrm{s}^\mathrm{R}
= \sum_{j=1}^L & \, \Big[ J_1 \hat{c}_{j \uparrow}^\dagger \hat{c}_{j \downarrow}
+ J_2 \hat{c}_{j \downarrow}^\dagger \hat{c}_{j + 1, \uparrow}
\\
& \  + \frac{\ii J_3}{2}
\big(
\hat{c}_{j \uparrow}^\dagger \hat{c}_{j + 1, \uparrow}
- \hat{c}_{j \downarrow}^\dagger \hat{c}_{j + 1, \downarrow}
\big)
+ \mathrm{H.c.} \Big],
\end{aligned}
\end{equation}
$\hat{L}_j^\mathrm{R} = \sqrt{\gamma} \, \hat{c}_{j \downarrow}^\dagger \hat{c}_{j \uparrow}$,
and $L$ denotes the system size.
The corresponding non-Hermitian effective Hamiltonian is
\begin{equation}
\begin{aligned}
\hat{H}_\mathrm{eff}^\mathrm{R}
= \hat{H}_\mathrm{s}^\mathrm{R}
- \frac{\ii \gamma}{2}
\sum_{j=1}^L \hat{n}_{j \uparrow} (1 - \hat{n}_{j \downarrow}),
\end{aligned}
\end{equation}
which reduces to the NHRM model if we omit the dissipative interaction term
$\frac{\ii \gamma}{2} \hat{n}_{j \uparrow} \hat{n}_{j \downarrow}$.

The quadratic Hermitian Hamiltonian $\hat{H}_\mathrm{s}^\mathrm{R}$ leads to opposite motion for $\uparrow$ and $\downarrow$ components.
To see this, we calculate the right- and left-moving amplitudes,
\begin{equation}
\begin{aligned}
g_{\pm}^{\nu} (t) &=
\langle j \pm 1,\nu | e^{-\ii \hat{H}_\mathrm{s}^\mathrm{R} t} | j \nu \rangle
\\&= \mp \xi_{\nu} \frac{J_3}{2} t -\frac{J_1 J_2}{2} t^2 + O(t^3),
\end{aligned}
\end{equation}
where $\nu \in \{ \uparrow, \downarrow \}$, $\xi_{\uparrow} = 1$ and $\xi_{\downarrow} = -1$.
Therefore,
\begin{equation}
|g_{+}^{\nu} (t)|^2 - |g_{-}^{\nu} (t)|^2
= \xi_{\nu} J_1 J_2 J_3 t^3 + O(t^4).
\end{equation}
We assume $J_1 J_2 J_3 > 0$, and thus the $\uparrow$ ($\downarrow$) component moves to the right (left).
The interacting non-Hermitian term $- \frac{\ii \gamma}{2} \sum_{j=1}^L \hat{n}_{j \uparrow} (1 - \hat{n}_{j \downarrow})$
induces loss for the $\uparrow$ component, 
leading to unidirectional propagation of quasiparticles to the left, 
which could be captured by the topology of complex quasiparticle spectra.

\begin{figure}[t]
\centering
\includegraphics[width=\columnwidth]{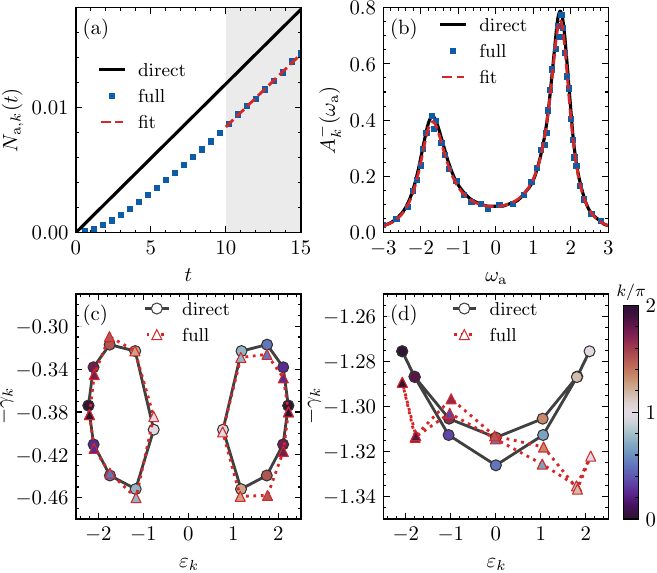}
\caption{(a) Auxiliary population $N_{\mathrm{a}, k} (t)$ as a function of $t$ for $k = \pi / 2$ and $\omega_\mathrm{a} = 1.8$.
The black line is directly calculated from Eq.~\eqref{Naktempty}, 
the blue squares are obtained from the simulation of the full Liouvillian that includes the auxiliary levels, 
and the red dashed line is a linear fit to the full simulation results over $t \in [10, 15]$.
(b) Spectral function $A_k^- (\omega_\mathrm{a})$ for $k = \pi / 2$.
The black line is calculated from Eqs.~\eqref{chimnst} and~\eqref{Akpm}, 
the blue squares are extracted by linearly fitting $N_{\mathrm{a}, k} (t)$ 
over $t \in [10, 15]$ for each $\omega_\mathrm{a}$, 
and the red dashed line is a fit to the full simulation results using Eq.~\eqref{fiteq}.
(c, d) Complex coherent quasiparticle spectra of the LRM model (c) and the AJ model (d) obtained by 
fitting the spectral functions from the direct calculations (solid lines with circles) and full simulations (dotted lines with triangles).
The momentum $k$ of each point is indicated by color.
In (a)--(c), we use the LRM model with $L = 8$, $N = 4$, $J_1 = 0.75$, $J_2 = 1.5$, $J_3 = 0.5$, $\gamma = 1.5$, and $\Omega = 0.08$.
In (d), we use the AJ model with $L = 12$, $N = 4$, $J_h = 1.0$, $\gamma_\mathrm{R} = 2.0$, $\gamma_\mathrm{L} = 0.4$, and $\Omega = 0.05$.
Here and henceforth, we consider periodic boundary conditions.
The direct calculations are performed in the doubled Hilbert space, 
while the full simulations use quantum jump evolution~\cite{Dalibard1992PRL}
averaged over 16384 samples for (a)--(c) and 65536 samples for (d).
}
\label{fig1}
\end{figure}

To measure the quasiparticle spectrum, we use the coupling Hamiltonian in Eq.~\eqref{couplingHamiltonian} with $m = \uparrow$.
In the numerical calculation, we initialize the auxiliary levels as empty ($N_{\mathrm{a}, k}^0 = 0$) 
and let particles transfer from the system to the auxiliary levels under the weak Rabi drive.
Then we measure the population $N_{\mathrm{a}, k} (t)$, which can be obtained from Eq.~\eqref{dotNakt} as
\begin{equation}\label{Naktempty}
N_{\mathrm{a}, k} (t) \approx \frac{\Omega^2 t}{4} A_k^- (\omega_\mathrm{a}).
\end{equation}
We consider a system with $N$ particles and perform calculations within the corresponding $N$-particle sector.
The initial system density matrix $\rho_\mathrm{s}$ is taken as the steady state of the system Liouvillian restricted to this sector.

Figure~\ref{fig1}(a) displays the auxiliary population $N_{\mathrm{a},k}(t)$ obtained 
from a full Liouvillian simulation where auxiliary levels are included
and a direct calculation based on Eq.~\eqref{Naktempty} using only the system Liouvillian.
At early times, the full simulation results deviate from the predicted linear behavior, 
whereas at late times the slope agrees well with the direct calculation.
By linearly fitting the simulation results at late times, 
we can extract $A_k^- (\omega_\mathrm{a})$ for all auxiliary frequencies $\omega_\mathrm{a}$.
The extracted spectral function is consistent with the direct calculation, as shown in Fig.~\ref{fig1}(b).
We then fit the spectral function using Eq.~\eqref{fiteq} to obtain
the real parts $\varepsilon_{k \alpha}$ and imaginary parts $-\gamma_{k \alpha}$ of the complex quasiparticle energies.
The fitted complex quasiparticle spectrum is shown in Fig.~\ref{fig1}(c), 
revealing the spectral topology associated with the unidirectional motion of quasiparticles.

We define the winding number for each branch $\alpha$ as~\cite{Gong2018PRX}
\begin{equation}
w_\alpha
= \frac{1}{2\pi}
\int_0^{2\pi} d k \,
\partial_k \arg [z_{k \alpha} - z_{\mathrm{ref}, \alpha}],
\end{equation}
where $z_{\mathrm{ref}, \alpha}$ is chosen inside the loop.
It follows from the Fourier transform 
$\hat{c}_{j m} = \frac{1}{\sqrt{L}} \sum_k e^{\ii k j} \hat{c}_{k m}$ that
a left-biased hopping contributes an $e^{\ii k}$ term to the complex spectrum,
giving rise to a counterclockwise winding as $k$ increases from $0$ to $2\pi$.
Indeed, for both branches in Fig.~\ref{fig1}(c), we find $w_\alpha = +1$,
consistent with left-moving quasiparticles.

To show that the protocol is not specific to the LRM model, 
we also apply it to the asymmetric-jump (AJ) model
that exhibits the Liouvillian skin effect~\cite{Haga2021PRL}.
The system Liouvillian for the AJ model (labeled by a superscript $\mathrm{A}$) is given by
\begin{equation}
\begin{aligned}
\mathcal{L}_\mathrm{s}^\mathrm{A} [\rho]
=& -\ii [\hat{H}_\mathrm{s}^\mathrm{A}, \rho]
\\&+ \sum_{j=1}^L \sum_{\mu = \mathrm{L}, \mathrm{R}}
\Big(
\hat{L}_{j \mu}^\mathrm{A} \rho  (\hat{L}_{j \mu}^\mathrm{A})^\dagger
- \frac{1}{2} \{ (\hat{L}_{j \mu}^\mathrm{A})^\dagger \hat{L}_{j \mu}^\mathrm{A}, \rho \}
\Big),
\end{aligned}
\end{equation}
where
\begin{equation}
\hat{H}_\mathrm{s}^\mathrm{A}
= -J_h \sum_{j=1}^L
(\hat{c}_{j+1}^\dagger \hat{c}_{j} + \mathrm{H.c.}),
\end{equation}
and
\begin{equation}
\hat{L}_{j\mathrm{L}}^\mathrm{A}
= \sqrt{\gamma_\mathrm{L}} \, \hat{c}_{j}^\dagger \hat{c}_{j+1},
\qquad
\hat{L}_{j\mathrm{R}}^\mathrm{A}
= \sqrt{\gamma_\mathrm{R}} \, \hat{c}_{j+1}^\dagger \hat{c}_{j}.
\end{equation}
Here $\gamma_\mathrm{R}$ and $\gamma_\mathrm{L}$ are the strengths for the right- and left-moving quantum jumps, respectively.

Applying the same protocol as in the LRM model, we obtain the quasiparticle spectrum shown in Fig.~\ref{fig1}(d).
The direct calculation yields a small loop in the complex-energy plane, 
which is well reproduced by the full simulation.
To quantify this agreement, we calculate the mean relative error
\begin{equation}
\delta x
= \frac{1}{N_x}
\sum_{n=1}^{N_x}
\frac{|x_n^\mathrm{full} - x_n^\mathrm{direct}|}{|x_n^\mathrm{direct}|},
\end{equation}
where we have excluded terms that have near-zero denominators (i.e., $\varepsilon_k$ at $k=\pi/2$ and $k=3\pi/2$).
For Fig.~\ref{fig1}(d), the mean relative errors are $\delta \varepsilon \approx 1.58 \% $, $\delta \gamma \approx 1.73 \%$, 
and $\delta z \approx 1.74 \% $.

The fitted spectrum in Fig.~\ref{fig1}(d) has winding number $w = +1$, 
corresponding to left-moving coherent quasiparticles,
although the jump rates satisfy $\gamma_\mathrm{R} > \gamma_\mathrm{L}$.
This can be understood by noting that the jump rates characterize 
incoherent processes and result in unidirectional transport in the diagonal sector,
whereas the coherent probe realized by the coupling Hamiltonian $\hat{H}_\mathrm{c}$
measures properties in the off-diagonal sector.

The weakly left-moving coherent quasiparticles can be understood from the real-space response
\begin{equation}
\chi_{r}^- (t) = -\ii \Tr \!
\big(
\hat{c}_{j + r}
e^{\mathcal{L}_\mathrm{s}^\mathrm{A} t}
[ \rho_\mathrm{s} \hat{c}_j^\dagger ]
\big),
\end{equation}
with $r = \pm 1$ and $t > 0$. 
The response is independent of $j$ due to translation symmetry.
Physically, $|\chi_{r}^- (t)|^2$ characterizes the propagation of a perturbation from $j$ to $j+r$ over time $t$.
At short times, by expanding
$e^{\mathcal{L}_\mathrm{s}^\mathrm{A} t} = \sum_{n=0}^\infty (\mathcal{L}_\mathrm{s}^\mathrm{A} t)^n / n!$
and numerically calculating the imbalance $|\chi_{+1}^- (t)|^2 - |\chi_{-1}^- (t)|^2$,
we find that the coefficients up to the fourth order vanish within numerical precision,
and thus the leading nonzero contribution is the fifth-order term
\begin{equation}
|\chi_{+1}^- (t)|^2 - |\chi_{-1}^- (t)|^2 = \kappa t^5 + O(t^6).
\end{equation}
For the parameters used in Fig.~\ref{fig1}(d), we have $\kappa \approx -2.30 \times 10^{-3}$,
which is consistent with weakly left-moving coherent quasiparticles extracted from the fitted spectrum.

\section{Incoherent Quasiparticles}
\label{sec_incoherent_quasiparticles}

The failure of coherent quasiparticle spectroscopy to capture the quantum jump induced transport in the AJ model
suggests the existence of quasiparticles intrinsic to open quantum systems, which lie in the incoherent (diagonal) sector.
To detect these quasiparticles, we define the response function as
\begin{equation}\label{inc_response}
X_k (t)
= -\ii \Theta (t) \Tr \! \big(
\mathcal{O}_k
e^{\mathcal{L}_\mathrm{s} t}
\mathcal{O}_k^\dagger [\rho_\mathrm{s}]
\big),
\end{equation}
where $\mathcal{O}_k$ is a superoperator acting on the density matrix,
with its adjoint denoted by $\mathcal{O}_k^\dagger$.
By choosing $\mathcal{O}_k[\rho] = \hat{c}_{k m} \rho$ and
$\mathcal{O}_k[\rho] = \rho \hat{c}_{k m}$,
$X_k (t)$ reduces to $\chi_k^+ (t)$ and $\chi_k^- (t)$, respectively.
For the incoherent quasiparticles, we choose $\mathcal{O}_k^\dagger = \mathcal{C}_{-k, m}$ with
\begin{equation}
\mathcal{C}_{k m}
= \frac{1}{\sqrt{L}} \sum_j e^{-\ii k j} \mathcal{C}_{j m},
\qquad
\mathcal{C}_{j m} [\rho]
= \hat{c}_{j m} \rho \hat{c}_{j m}^\dagger,
\end{equation}
which acts on both sides of the density matrix so that the response probes the incoherent sector.
The associated spectral function is given by
\begin{equation}\label{inc_spectral}
B_k (\omega)
= -2 \operatorname{Im}
\int_0^\infty d t \, e^{\ii \omega t} X_k (t).
\end{equation}

The choice $\mathcal{O}_k^\dagger = \mathcal{C}_{-k, m}$ ensures that
the perturbation $\mathcal{C}_{-k, m} [\rho_\mathrm{s}]$ carries momentum $k$,
since $\hat{T} \mathcal{C}_{-k, m} [\rho_\mathrm{s}] \hat{T}^{-1}
= e^{-\ii k} \mathcal{C}_{-k, m} [\rho_\mathrm{s}]$,
with $\hat{T}$ being the translation operator satisfying
$\hat{T} \hat{c}_{j m} \hat{T}^{-1} = \hat{c}_{j+1, m}$ and 
$\hat{T} \rho_\mathrm{s} \hat{T}^{-1} = \rho_\mathrm{s}$.
This is consistent with the perturbations 
$\hat{c}_{k m}^\dagger \rho_\mathrm{s}$ and $\rho_\mathrm{s} \hat{c}_{k m}^\dagger$
in the coherent response functions $\chi_k^+ (t)$ and $\chi_k^- (t)$, respectively, since
$\hat{T} \hat{c}_{k m}^\dagger \rho_\mathrm{s} \hat{T}^{-1}
= e^{-\ii k} \hat{c}_{k m}^\dagger \rho_\mathrm{s}$ and
$\hat{T} \rho_\mathrm{s} \hat{c}_{k m}^\dagger \hat{T}^{-1}
= e^{-\ii k} \rho_\mathrm{s} \hat{c}_{k m}^\dagger$.

To measure the spectral function $B_k (\omega)$, 
we prepare empty auxiliary levels and apply a short dissipative pulse 
described by $e^{\mathcal{L}_\mathrm{d} t_\mathrm{d}}$, where
\begin{equation}\label{dissipative_L_d}
\mathcal{L}_\mathrm{d} [\rho]
=
\sum_j
\Big(
\hat{L}_{\mathrm{d}, j} \rho \hat{L}_{\mathrm{d}, j}^\dagger
-
\frac{1}{2}
\{ \hat{L}_{\mathrm{d}, j}^\dagger \hat{L}_{\mathrm{d}, j}, \rho \}
\Big),
\end{equation}
with jump operators
\begin{equation}
\hat{L}_{\mathrm{d}, j} =
\sqrt{\gamma_\mathrm{d} [1+\cos(k j)]} \,
\hat{a}_j^\dagger \hat{c}_{j m}.
\end{equation}
Here $\gamma_\mathrm{d}$ is large so that the system Liouvillian can be neglected during the pulse,
while we take the pulse duration $t_\mathrm{d}$ to be short to keep the response approximately linear.
Then we turn off the dissipative pulse and perform Liouvillian time evolution under $\mathcal{L}_\mathrm{s}$ for time $t$, 
which gives
\begin{equation}
\label{incrhot1}
\rho (t) = e^{\mathcal{L}_\mathrm{s} t}
e^{\mathcal{L}_\mathrm{d} t_\mathrm{d}} [\rho_0],
\end{equation}
where the initial state is given by $\rho_0 = \rho_\mathrm{s} \otimes |0\rangle_\mathrm{a} \langle 0|_\mathrm{a}$.
After the evolution, we measure the operator $\hat{M}_{k,n} = \hat{N}_\mathrm{a} \hat{D}_{k,n}$ and obtain
\begin{equation}
\label{inc_measurement}
M_{k, n} (t) = \Tr \! \big[ \hat{M}_{k,n} \rho (t) \big],
\end{equation}
where 
\begin{align}
\label{inc_aux_number}
\hat{N}_\mathrm{a} &= \sum_j \hat{a}_j^\dagger \hat{a}_j,
\\
\label{inc_density_readout}
\hat{D}_{k,n} &= \sum_j \cos \! \Big( k j - \frac{n \pi}{2} \Big) \, \hat{c}_{j m} \hat{c}_{j m}^\dagger.
\end{align}
In Appendix~\ref{inc_experiment}, we discuss how to engineer $\mathcal{L}_\mathrm{d}$ and measure $M_{k,n}(t)$ experimentally.

Using linear response theory, we obtain
\begin{equation}\label{inc_lrt}
\begin{aligned}
M_{k,0}(t) &= -\frac{\Theta_\mathrm{d} L}{2}
(1 + \delta_{k\pi} + 3\delta_{k0})
\operatorname{Im} X_k(t) + O(\Theta_\mathrm{d}^2),
\\
M_{k,1}(t) &= -\frac{\Theta_\mathrm{d} L}{2}
(1 + \delta_{k\pi} + 3\delta_{k0})
\operatorname{Re} X_k(t) + O(\Theta_\mathrm{d}^2),
\end{aligned}
\end{equation}
where $\Theta_\mathrm{d} = \gamma_\mathrm{d} t_\mathrm{d}$.
The derivation is given in Appendix~\ref{inc_derivation}.
Therefore, the spectral function of the incoherent quasiparticles can be obtained as
\begin{equation}\label{inc_spectral_aux}
\begin{aligned}
B_{k} (\omega) \approx \,
&\frac{4}{\Theta_\mathrm{d} L (1 + \delta_{k\pi} + 3\delta_{k0})}
\\ &\times
\int_0^\infty dt \,
[ \cos ( \omega t ) M_{k,0} (t) + \sin ( \omega t ) M_{k,1} (t) ].
\end{aligned}
\end{equation}

We now demonstrate the protocol using concrete models.
For clarity, we first consider the LRM model in a parameter regime where the incoherent quasiparticle spectrum can be obtained analytically.
We take $\rho_\mathrm{s}$ to be the steady state in the $N=2$ particle-number sector and choose $m=\uparrow$.
We further set $J_1 = J_3 = 0$, such that the Hamiltonian only contains the hopping between $|j \downarrow \rangle$ and $|j+1, \uparrow \rangle$, 
while the jump operator transfers $|j \uparrow \rangle$ to $|j \downarrow \rangle$.
Thus, the particle transfer follows the sequence
$|j\uparrow \rangle \rightarrow |j\downarrow \rangle
\leftrightarrow |j+1,\uparrow \rangle \rightarrow |j+1, \downarrow \rangle$, 
leading to unidirectional particle propagation to the right.

\begin{figure}[t]
\centering
\includegraphics[width=\columnwidth]{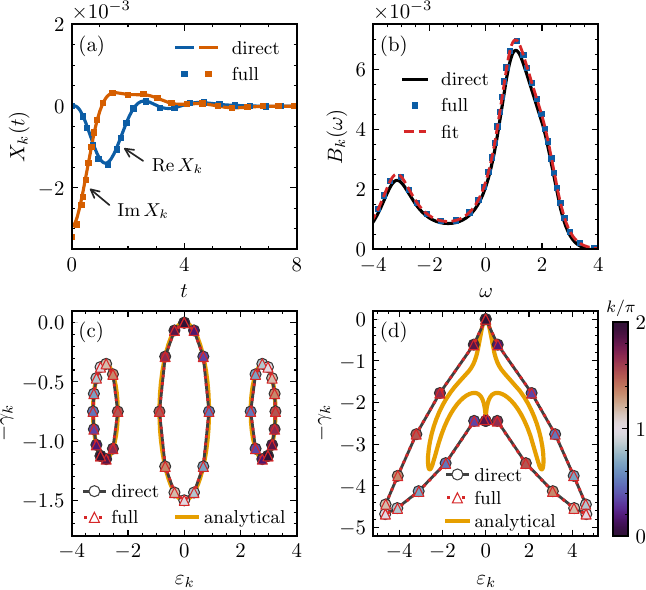}
\caption{(a) Real and imaginary parts of $X_k(t)$ for $k=\pi/2$.
The solid lines are calculated from the definition of $X_k(t)$ [Eq.~\eqref{inc_response}], 
and the squares are obtained from a full simulation of the protocol using Eq.~\eqref{inc_lrt}.
(b) Spectral function $B_k(\omega)$ for $k=\pi/2$.
The black line and blue squares are obtained from a direct calculation 
using Eq.~\eqref{inc_spectral} and a full simulation using Eq.~\eqref{inc_spectral_aux}, respectively, 
and the red dashed line is a fit to the full simulation results using Eq.~\eqref{fiteq}.
(c, d) Complex incoherent quasiparticle spectra of the LRM model (c) and the AJ model (d)
obtained by fitting the direct calculation (solid lines with circles) and full simulation results (dotted lines with triangles).
The $k=0$ points are treated separately in the fitting procedure, as detailed in Appendix~\ref{inc_k0}.
The momentum $k$ of each point is indicated by color.
The yellow lines in (c) and (d) display the eigenvalues of
$H_\mathrm{inc}^\mathrm{R}(k)$ [Eq.~\eqref{inc_modelA_k}] and
$\tilde{H}_\mathrm{inc}^\mathrm{A}(k)$ [Eq.~\eqref{inc_modelB_TB}], respectively.
In (a)--(c), we use the LRM model with $L=12$, $N=2$, $J_1=J_3=0$, $J_2=1.5$, $\gamma=1.5$, and $\Theta_\mathrm{d}=10^{-2}$.
In (d), we use the AJ model with $L=12$, $N=4$, $J_h=1.0$, $\gamma_\mathrm{R}=2.0$, $\gamma_\mathrm{L}=0.4$, and $\Theta_\mathrm{d}=10^{-2}$.
The direct calculations and full simulations are performed in the doubled Hilbert space.
For the full simulations we use the dynamics of the reduced density matrix detailed in Appendix~\ref{appendix1}.
}
\label{fig2}
\end{figure}

As shown in Figs.~\ref{fig2}(a) and~\ref{fig2}(b), 
the full simulation results accurately reproduce 
both $X_k (t)$ and $B_k (\omega)$ obtained from direct calculations.
We further fit the spectral functions at each momentum $k$ to obtain the complex quasiparticle spectrum shown in Fig.~\ref{fig2}(c).
The quasiparticle energies extracted from the full simulation results agree well with those obtained from the directly calculated spectral functions.
Each branch possesses a winding number $w_\alpha = -1$ around a reference point inside the corresponding loop, 
consistent with the unidirectional propagation to the right.

We now provide an analytical description of the incoherent quasiparticle spectrum.
We first note that the perturbation $\mathcal{C}_{-k,\uparrow}[\rho_\mathrm{s}]$
in $X_k(t)$ evolves in the single-particle sector, given that 
$\mathcal{C}_{-k,\uparrow}$ removes one particle from the $N=2$ steady state $\rho_\mathrm{s}$
and $\mathcal{L}_\mathrm{s}$ conserves the particle number.
Thus, the response is determined by the time-dependent single-particle density matrix elements.
For a density matrix $\rho$ in the single-particle sector, we define
\begin{equation}
\begin{aligned}
u_j &= \langle j  \uparrow | \rho | j \uparrow \rangle,
\\
d_j &= \langle j \downarrow | \rho | j \downarrow \rangle,
\\
y_j & = \frac{\ii}{2} \big( 
\langle j+1, \uparrow | \rho | j \downarrow \rangle 
- \langle j \downarrow | \rho |j+1, \uparrow \rangle 
\big).
\end{aligned}
\end{equation}
For $J_1 = J_3 = 0$, the master equation gives
\begin{equation}\label{inc_modelA_real}
\begin{aligned}
\dot{u}_j &= -\gamma u_j + 2 J_2 y_{j-1},
\\
\dot{d}_j &= \gamma u_j - 2 J_2 y_j,
\\
\dot{y}_j &= - J_2 u_{j+1} + J_2 d_j - \frac{\gamma}{2} y_j.
\end{aligned}
\end{equation}
Applying the Fourier transform $u_k = \frac{1}{\sqrt{L}} \sum_j e^{-\ii k j} u_j$, 
with similar definitions for $d_k$ and $y_k$, we obtain
\begin{align}
\frac{d}{dt}
\begin{pmatrix}
u_k\\ d_k\\ y_k
\end{pmatrix}
&=
- \ii H_\mathrm{inc}^\mathrm{R} (k)
\begin{pmatrix}
u_k\\ d_k\\ y_k
\end{pmatrix},
\\
\label{inc_modelA_k}
H_\mathrm{inc}^\mathrm{R} (k)
&= \ii
\begin{pmatrix}
-\gamma & 0 & 2J_2e^{-\ii k}
\\
\gamma & 0 & -2J_2
\\
-J_2e^{\ii k} & J_2 & -\gamma/2
\end{pmatrix}.
\end{align}

We plot the eigenvalues of $H_\mathrm{inc}^\mathrm{R}(k)$ 
with respect to $k$ in Fig.~\ref{fig2}(c),
which agree quantitatively with the quasiparticle energies extracted from the direct calculations 
and full simulations at each $k$.
Specifically, the mean relative error between the nonzero complex energies extracted from direct calculations and the analytical eigenvalues is below $10^{-5}$,
showing that the measured incoherent quasiparticles indeed originate from single-particle physics in the incoherent sector.

We next consider the AJ model with the steady state $\rho_\mathrm{s}$ 
in the $N=4$ particle-number sector, where many-body effects are present.
Following the same procedure, we extract the quasiparticle spectrum shown in Fig.~\ref{fig2}(d)
and find excellent agreement between the direct calculations and full simulations.
The two branches connect to form a single loop with a winding number of $w=-1$,
which is consistent with the right-moving incoherent quasiparticles for
$\gamma_\mathrm{R} > \gamma_\mathrm{L}$.

To obtain an analytical description of the incoherent quasiparticle spectrum in the many-body case,
we employ a fermionic Gaussian state as a variational ansatz to derive an effective Hamiltonian 
governing the dynamics of the correlation function (see Appendix~\ref{appendix2}),
\begin{equation}\label{inc_modelB_TB}
\tilde{H}_\mathrm{inc}^\mathrm{A} (k)
= \begin{pmatrix} \ii \tilde{h}(k) & \sqrt{2} J_h (e^{-\ii k} - 1)
\\
\sqrt{2} J_h (e^{\ii k} - 1) & -\ii (\gamma_\mathrm{R} + \gamma_\mathrm{L})
\end{pmatrix},
\end{equation}
where
\begin{equation}
\tilde{h}(k)
= \tilde{\gamma}_\mathrm{R} (e^{\ii k} - 1)
+ \tilde{\gamma}_\mathrm{L} (e^{-\ii k} - 1),
\end{equation}
with
$\tilde{\gamma}_\mathrm{R}
= \gamma_\mathrm{R}
- (1 - \bar{n})(\gamma_\mathrm{R}-\gamma_\mathrm{L})$,
$\tilde{\gamma}_\mathrm{L}
= \gamma_\mathrm{L}
+ (1 - \bar{n})(\gamma_\mathrm{R}-\gamma_\mathrm{L})$,
and $\bar{n} = N/L$ being the filling fraction.
The eigenvalues of $\tilde{H}_\mathrm{inc}^\mathrm{A}(k)$ are plotted as a yellow line 
in Fig.~\ref{fig2}(d).
The resulting loop exhibits a winding number of $w=-1$ 
around a reference point located inside both the analytical and numerical loops,
thereby capturing the nontrivial spectral winding induced by the asymmetric jump processes.

\section{Quasiparticle Spectroscopy for Spontaneous Symmetry Breaking}

We further apply the spectroscopy protocols developed above to detect SSB 
in open quantum systems, where symmetries can be either strong or weak,
giving rise to richer forms of SSB than in closed systems.
Using models with weak and strong $U(1)$ symmetries, we demonstrate 
how symmetry breaking manifests in the coherent and incoherent quasiparticle spectra.

\subsection{Weak $U(1)$ symmetry}
\label{subsec_weak_u1}

We consider a one-dimensional hard-core boson model adapted from Ref.~\cite{Gu2025PRB}, 
where the lattice consists of $L$ unit cells with two sublattices $A$ and $B$.
The system Liouvillian, designated by a superscript $\mathrm{W}$, is given by
\begin{equation}\label{weak_u1_liouvillian}
\begin{aligned}
\mathcal{L}_\mathrm{s}^\mathrm{W}[\rho]
= & -\ii[\hat{H}_\mathrm{s}^\mathrm{W},\rho]
\\
& + \sum_{j = 1}^L \sum_{\mu = A, B}
\Big(
\hat{L}_{j\mu}^\mathrm{W}\rho(\hat{L}_{j\mu}^\mathrm{W})^\dagger
-\frac{1}{2}\{(\hat{L}_{j\mu}^\mathrm{W})^\dagger\hat{L}_{j\mu}^\mathrm{W},\rho\}
\Big),
\end{aligned}
\end{equation}
with Hamiltonian
\begin{equation}\label{weak_u1_hamiltonian}
\hat{H}_\mathrm{s}^\mathrm{W}
= J \sum_{ij}
\frac{\hat{b}_{iA}^\dagger\hat{b}_{jB}+\hat{b}_{jB}^\dagger\hat{b}_{iA}}{|2(j-i)+1|^{p}}
\end{equation}
and jump operators
\begin{equation}\label{weak_u1_jumps}
\hat{L}_{jA}^\mathrm{W}=\sqrt{\gamma}\,\hat{b}_{jA}^\dagger,
\qquad
\hat{L}_{jB}^\mathrm{W}=\sqrt{\gamma}\,\hat{b}_{jB}.
\end{equation}
Here $\hat{b}_{j\nu}$ annihilates a hard-core boson on sublattice $\nu$ in unit cell $j$,
$J$ is the hopping strength, and $\gamma$ is the local gain rate for A sites and loss rate for B sites.
We extend Model~I of Ref.~\cite{Gu2025PRB}
by replacing its nearest-neighbor Hamiltonian with the long-range Hamiltonian in Eq.~\eqref{weak_u1_hamiltonian},
where the exponent $p$ governs the hopping range.
The Liouvillian possesses a weak $U(1)$ symmetry since
\begin{equation}\label{weak_u1_covariance}
\hat{U}_\theta \mathcal{L}_\mathrm{s}^\mathrm{W} [\rho] \hat{U}_\theta^\dagger
= \mathcal{L}_\mathrm{s}^\mathrm{W} [\hat{U}_\theta \rho \hat{U}_\theta^\dagger],
\end{equation}
where $\hat{U}_\theta = e^{\ii\theta\hat{N}_{\mathrm{s}}}$ with 
$\hat{N}_{\mathrm{s}}=\sum_{j \nu}\hat{b}_{j \nu}^\dagger\hat{b}_{j \nu}$ 
being the total particle number operator for the system levels.
On the other hand, $[\hat{U}_\theta, \hat{L}_{j \mu}^\mathrm{W}]\neq 0$ for generic $\theta$ 
means that the Liouvillian does not possess a strong $U(1)$ symmetry.

\begin{figure}[t]
\centering
\includegraphics[width=\columnwidth]{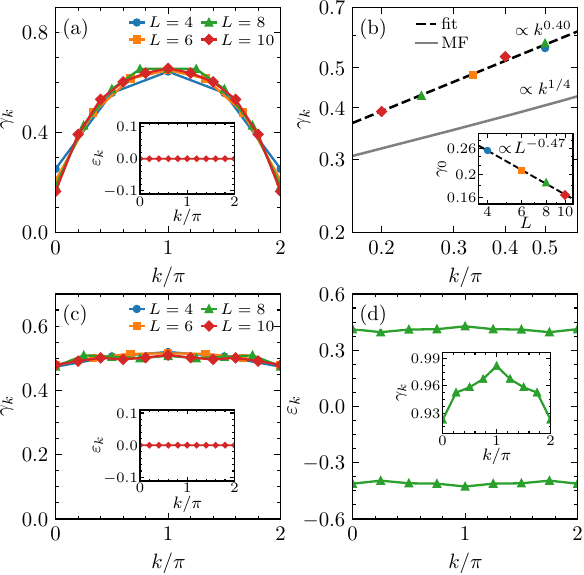}
\caption{(a) Decay rates $\gamma_k$ of coherent quasiparticles 
with respect to $k$ for different system sizes $L$ in the symmetry-broken regime.
The inset shows the real parts $\varepsilon_k$ of the coherent quasiparticle energies for $L = 10$.
(b) Scaling of $\gamma_k$ in (a) at small $k$ on a log-log scale.
The black dashed line is a power-law fit and the gray solid line is the mean-field result from Eq.~\eqref{wu1_eig}.
The inset shows $\gamma_0$ as a function of $L$ on a log-log scale, where the dashed line denotes a power-law fit.
(c) Same as (a), but in the symmetry-unbroken regime.
(d) Real parts $\varepsilon_k$ of the incoherent quasiparticle energies
with respect to $k$ for $L = 8$ in the symmetry-broken regime.
The inset shows $\gamma_k$ for both branches, which approximately coincide.
We choose $J = 2 J_\mathrm{c}^{\mathrm{MF}}$ in (a), (b), and (d),
and $J = 0.4 J_\mathrm{c}^{\mathrm{MF}}$ in (c).
In all panels, we take $\gamma = 1$ and probe sublattice $A$.
The quantum jump evolution is averaged over 4096 samples for (a)--(c)
and 65536 samples for (d).}
\label{fig3}
\end{figure}

A mean-field analysis (see Appendix~\ref{appendix_weak_u1_mf}) predicts two steady-state phases
with order parameters being $\psi_{j \nu} = \langle \hat{b}_{j \nu} \rangle$.
Given the mean-field critical point
\begin{equation}\label{weak_u1_mf_boundary}
J_\mathrm{c}^{\mathrm{MF}} = \frac{\gamma}{4 (1-2^{-p}) \zeta(p)},
\end{equation}
where $\zeta(p)$ is the Riemann zeta function, 
the two phases are characterized by
\begin{equation}\label{weak_u1_mf_order}
|\psi_{A}|^2 = |\psi_{B}|^2 =
\begin{cases}
0, & J < J_\mathrm{c}^{\mathrm{MF}},\\
\eta (1-\eta) / 2, & J > J_\mathrm{c}^{\mathrm{MF}},
\end{cases}
\end{equation}
where $\psi_{j \nu} = \psi_{\nu}$ due to translation symmetry, and $\eta = J_\mathrm{c}^{\mathrm{MF}}/J$.
For $J < J_\mathrm{c}^{\mathrm{MF}}$, the weak $U(1)$ symmetry is preserved.
In contrast, for $J > J_\mathrm{c}^{\mathrm{MF}}$, spontaneous breaking of the weak $U(1)$ symmetry occurs, 
resulting in a nonzero $\psi_{j \nu}$.
Further analysis suggests that long-range order can survive
if the interaction is sufficiently long-ranged
(e.g., for $p = 5/4$; see Appendix~\ref{wu1_cluster}).
We therefore adopt $p = 5/4$ for the subsequent calculations.

Using quantum jump evolution, we directly evaluate the coherent response function
$\chi_k^- (t)$ in Eq.~\eqref{chimnst},
with $\hat{c}_{j m}$ replaced by $\hat{b}_{j m}$ for $m = A$.
We then obtain the corresponding spectral function $A_k^- (\omega)$ using Eq.~\eqref{Akpm}
and fit it to Eq.~\eqref{fiteq} to extract the complex quasiparticle energies.
In the weak $U(1)$ symmetry-broken regime,
the quasiparticle decay rates $\gamma_k$ approach zero at $k = 0$ as $L$ increases,
and the real part $\varepsilon_k$ vanishes, as illustrated in Fig.~\ref{fig3}(a).
This behavior signals a gapless coherent quasiparticle spectrum associated with
the spontaneous breaking of the continuous weak $U(1)$ symmetry.

To further investigate this gapless behavior, we fit $\gamma_k$ at small, nonzero $k$
to a power-law function, obtaining $\gamma_k \propto k^{0.40}$ as shown in Fig.~\ref{fig3}(b).
The numerical results agree well with the fit, 
supporting the presence of a gapless dispersion relation in the thermodynamic limit.
We also fit $\gamma_0$ to a power-law function of $L$ and obtain $\gamma_0 \propto L^{-0.47}$
[see the inset of Fig.~\ref{fig3}(b)].
The gapless spectrum arises because the spontaneous breaking
of the weak $U(1)$ symmetry gives rise to
Nambu--Goldstone modes~\cite{Sieberer2016RPP}.
These modes correspond to long-wavelength phase fluctuations of the order parameter, expressed as
$\psi_{j \nu}^{(k)} \simeq \psi_{\nu} e^{\ii \Delta \theta_{j \nu}^{(k)} }$, where the fluctuation
$\Delta \theta_{j \nu}^{(k)} = \Delta \theta e^{\ii k j}$ is sufficiently small.
Expanding this to linear order yields $\psi_{j \nu}^{(k)} - \psi_{\nu} \simeq  \ii \psi_{\nu} \Delta \theta e^{\ii k j}$.
The perturbation $\rho_\mathrm{s} \hat{b}_{k m}^\dagger$ 
in the response function $\chi_k^-(t)$ is expected to induce such fluctuations
as $ \Tr (\rho_\mathrm{s} \hat{b}_{k m}^\dagger \hat{b}_{j m} ) \sim e^{\ii k j}$.
Consequently, $\chi_k^-(t) \sim e^{-\ii z_k t}$ is dominated by the Nambu--Goldstone mode with complex energy $z_k$,
allowing us to detect spontaneous breaking of the weak $U(1)$ symmetry through coherent quasiparticle spectroscopy.
In Appendix~\ref{wu1_fluct}, we derive the mean-field dispersion relation of the Nambu--Goldstone mode,
which scales as $k^{1/4}$ at small $k$ for $p=5/4$ [see the gray line in Fig.~\ref{fig3}(b)].

In contrast, the coherent quasiparticle spectrum is gapped in the symmetry-unbroken regime.
As shown in Fig.~\ref{fig3}(c), the decay rates $\gamma_k$ remain large even as $L$ increases,
while the real parts $\varepsilon_k$ stay approximately zero.
This behavior is consistent with the fact that 
the weak $U(1)$ symmetry remains unbroken, preventing the emergence of a gapless Nambu--Goldstone mode.
We therefore conclude that coherent quasiparticle spectroscopy can reveal 
spontaneous breaking of the weak $U(1)$ symmetry by detecting gapless Nambu--Goldstone modes.

Finally, we apply the same procedure to the incoherent response function $X_k (t)$ 
[Eq.~\eqref{inc_response}] to obtain the incoherent quasiparticle spectrum in the symmetry-broken regime.
As shown in Fig.~\ref{fig3}(d), the spectrum features two branches with nonzero real parts $\varepsilon_k$
and approximately identical decay rates $\gamma_k$.
Both branches are gapped in this regime, 
in stark contrast to the gapless coherent quasiparticle spectrum in Fig.~\ref{fig3}(a).

\subsection{Strong $U(1)$ symmetry}
\label{subsec_strong_u1}

We next consider a Liouvillian with strong $U(1)$ symmetry adapted from Model~II of Ref.~\cite{Gu2025PRB},
\begin{equation}
\begin{aligned}
\mathcal{L}_\mathrm{s}^\mathrm{S} [\rho]
= {}& - \ii [\hat{H}_\mathrm{s}^\mathrm{S}, \rho]
\\
&+ \sum_{j = 1}^L \sum_{\mu = 1}^4
\Big(
\hat{L}_{j\mu}^\mathrm{S} \rho (\hat{L}_{j\mu}^\mathrm{S})^\dagger
- \frac{1}{2}
\{ (\hat{L}_{j\mu}^\mathrm{S})^\dagger \hat{L}_{j\mu}^\mathrm{S}, \rho \}
\Big),
\end{aligned}
\end{equation}
with Hamiltonian $\hat{H}_\mathrm{s}^\mathrm{S} = \hat{H}_\mathrm{s}^\mathrm{W}$
in Eq.~\eqref{weak_u1_hamiltonian} and jump operators
\begin{equation}
\begin{aligned}
\hat{L}_{j1}^\mathrm{S}
&= \sqrt{\gamma} \, \hat{b}_{jB}^\dagger \hat{b}_{jA},
&
\hat{L}_{j2}^\mathrm{S}
&= \sqrt{\gamma} \, \hat{b}_{j-1,B}^\dagger \hat{b}_{jA},
\\
\hat{L}_{j3}^\mathrm{S}
&= \sqrt{\gamma_z} \, \hat{\sigma}_{jA}^z,
&
\hat{L}_{j4}^\mathrm{S}
&= \sqrt{\gamma_z} \, \hat{\sigma}_{jB}^z,
\end{aligned}
\end{equation}
where $\hat{\sigma}_{j\nu}^z = 1 - 2 \hat{b}_{j\nu}^\dagger \hat{b}_{j\nu}$.
Here $\gamma$ is the jump rate from sublattice $A$
to its neighboring $B$ sites, and $\gamma_z$ is the local dephasing strength.
The Liouvillian has a strong $U(1)$ symmetry since
$[\hat{U}_\theta, \hat{H}_\mathrm{s}^\mathrm{S}] = 0$ and
$[\hat{U}_\theta, \hat{L}_{j\mu}^\mathrm{S}] = 0$,
with $\hat{U}_\theta = e^{\ii \theta \hat{N}_{\mathrm{s}}}$.
Therefore, the system can be restricted to a sector with a fixed particle number, and
we focus on the half-filled sector with particle number $N = L$.

The strong $U(1)$ symmetry of a density matrix $\rho$ can be expressed as
$\hat{U}_\theta \rho = e^{\ii \theta N} \rho$, where $N$ is the particle number.
Notably, the strong $U(1)$ symmetry also ensures the weak $U(1)$ symmetry,
$\hat{U}_\theta \rho \hat{U}_\theta^\dagger = \rho$.
Strong-to-weak $U(1)$ SSB occurs when the strong symmetry
is spontaneously broken while the weak symmetry remains unbroken.
This symmetry breaking can be further understood by vectorizing the density matrix as
$|\rho\rangle = \sum_{nn'} \rho_{n n'} |n\rangle \otimes |n'\rangle$,
where $\rho_{n n'} = \langle n | \rho | n' \rangle$.
In the doubled Hilbert space, strong $U(1)$ symmetry corresponds to 
two independent $U(1)$ rotations on the ket and bra spaces generated by 
$\hat{N}_L = \hat{N}_{\mathrm{s}} \otimes \hat{I}$ and $\hat{N}_R = \hat{I} \otimes \hat{N}_{\mathrm{s}}$, respectively.
These rotations can be recombined into
$\hat{U}_\theta^\mathrm{A} = e^{\ii \theta (\hat{N}_L + \hat{N}_R)/2}$ and
$\hat{U}_\theta^\mathrm{W} = e^{\ii \theta (\hat{N}_L - \hat{N}_R)}$. 
Here, $\hat{U}_\theta^\mathrm{W}$ acts as the weak $U(1)$ transformation
$\hat{U}_\theta^\mathrm{W} | \rho \rangle = | \hat{U}_\theta \rho \hat{U}_\theta^\dagger \rangle$.
Strong-to-weak $U(1)$ SSB then corresponds to the spontaneous breaking of the symmetry represented by 
$\hat{U}_\theta^\mathrm{A}$.

\begin{figure}[t]
\centering
\includegraphics[width=\columnwidth]{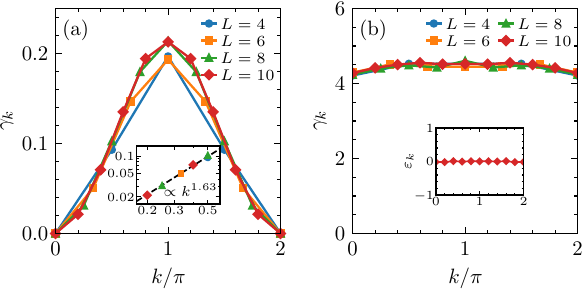}
\caption{(a) Decay rates $\gamma_k$ of incoherent quasiparticles as a function of $k$ 
for different system sizes $L$ in the regime featuring strong-to-weak $U(1)$ SSB.
The inset shows the scaling of $\gamma_k$ at small $k$ on a log-log scale,
where the black dashed line represents a power-law fit $\gamma_k \propto k^{1.63}$.
(b) Decay rates $\gamma_k$ of coherent quasiparticles for different system sizes $L$.
The inset shows the real parts $\varepsilon_k$ of the coherent quasiparticle energies for $L=10$.
Here, the parameters are chosen as $p=3/2$, $J = 0.592$, $\gamma=1$, $\gamma_z = 2$, and $N=L$, with sublattice $B$ being probed.
The data are obtained using quantum jump evolution averaged over 32768 samples in (a) and 4096 samples in (b).
}
\label{fig4}
\end{figure}

To probe strong-to-weak $U(1)$ SSB, we use incoherent quasiparticle spectroscopy,
with the results shown in Fig.~\ref{fig4}(a).
The parameters are chosen on the basis of the mean-field analysis in Appendix~\ref{su1_mf}.
The decay rates $\gamma_k$ decrease toward zero at small $k$ according to 
$\gamma_k \propto k^{1.63}$ [see the inset of Fig.~\ref{fig4}(a)],
and the real parts $\varepsilon_k$ stay approximately zero (not shown here),
suggesting a gapless incoherent quasiparticle spectrum.
At $k=0$, the strong $U(1)$ symmetry ensures a nonzero $X_0 (t)$ for $t \to \infty$,
leading to $\gamma_0 = 0$ (see Appendix~\ref{inc_k0}).
In contrast, the coherent spectrum remains gapped as $L$ increases [see Fig.~\ref{fig4}(b)],
consistent with the unbroken weak $U(1)$ symmetry.

The perturbation $|\mathcal{O}_k^\dagger[\rho_\mathrm{s}]\rangle$
is expected to generate Nambu--Goldstone modes associated with the
strong-to-weak $U(1)$ SSB~\cite{Gu2025PRB}.
This can be understood by noting that the operator $\hat{b}_{jm} \otimes \hat{b}_{jm}$,
which transforms nontrivially under the strong symmetry but remains invariant under the weak symmetry,
serves as an order parameter for this symmetry breaking (see Appendix~\ref{su1_mf}).
Since $\mathcal{O}_k^\dagger = \frac{1}{\sqrt{L}} \sum_j e^{\ii k j}
(\hat{b}_{jm} \otimes \hat{b}_{jm})$ in the doubled Hilbert space,
it creates long-wavelength fluctuations of this order parameter at small $k$, analogous to the weak $U(1)$ case.
The resulting response $X_k(t) \sim e^{-\ii z_k t}$ can then be used to extract the complex energy $z_k$
of the Nambu--Goldstone mode at small $k$,
whose gapless dispersion signals the strong-to-weak $U(1)$ SSB.

\section{Summary and Outlook}
\label{sec_summary_outlook}

In this work, we introduce the concept of coherent and incoherent Liouvillian quasiparticles
and propose spectroscopy protocols to measure their complex energies.
We demonstrate the feasibility of these protocols
through numerical simulations of the LRM and AJ models.
The complex spectra obtained from these protocols are in excellent agreement 
with those from direct calculations, revealing nontrivial spectral topology
associated with unidirectional quasiparticle transport.
Furthermore, we show how SSB in open quantum systems
manifests in the Liouvillian quasiparticle spectra.
For weak $U(1)$ SSB, 
our results unveil a gapless Nambu--Goldstone mode 
in the coherent spectrum alongside a gapped incoherent spectrum. 
For strong-to-weak $U(1)$ SSB, the incoherent spectrum 
becomes gapless while the coherent spectrum remains gapped.
The nature of gapless quasiparticles thus reflects 
the specific type of $U(1)$ SSB.

Our methods can be extended to study other exotic phenomena
in open quantum many-body systems.
For example, mixed states can exhibit nontrivial topology protected
by a combination of strong and weak symmetries~\cite{Ma2023PRX,Guo2025PRX,Ma2025PRX}.
A natural question that arises is how to characterize the boundary modes associated with 
these nontrivial states.
We expect that by extending our protocols to site-resolved spectroscopy, 
one can detect these boundary modes and measure their complex excitation energies.
Furthermore, coherent and incoherent quasiparticles can differ significantly in their
transport properties, even propagating in opposite directions
within the same system, as demonstrated in Secs.~\ref{sec_coherent_quasiparticles}
and~\ref{sec_incoherent_quasiparticles}.
These findings motivate further study of transport in 
open quantum systems from a quasiparticle perspective.
Another direction is to employ Liouvillian quasiparticle spectroscopy
to investigate exceptional points and their associated exotic features,
such as the braiding of complex energy bands~\cite{Hu2021PRL}
and bulk Fermi arcs~\cite{Kozii2024PRB}.
Thus, our protocols open a new avenue for understanding
open quantum matter through coherent and incoherent
Liouvillian quasiparticles.

\begin{acknowledgments}
We thank Hongchao Li and Ming Gong for helpful discussions.
M.U. was supported by the RIKEN TRIP initiative.
\end{acknowledgments}

\section*{Data availability}
The data and code for our numerical simulations are available
from the corresponding authors upon reasonable request.

\appendix

\section{Linear response theory for coherent quasiparticle spectroscopy}
\label{coh_derivation}

Here we derive Eq.~\eqref{dotNakt}
under the assumption that the coupling between the system and auxiliary levels is weak.
We first separate the coupling Hamiltonian into the detuning and Rabi terms,
\begin{align}
\hat{H}_\mathrm{c} &= \hat{H}_\mathrm{a} + \hat{V},
\\
\hat{H}_\mathrm{a} &= \omega_\mathrm{a} \sum_q \hat{a}_q^\dagger \hat{a}_q,
\\
\label{coh_rabi_coupling}
\hat{V} &= \frac{\Omega}{2} \sum_q 
\big( \hat{a}_q^\dagger \hat{c}_{q m} + \hat{c}_{q m}^\dagger \hat{a}_q \big).
\end{align}
Then we define the unperturbed Liouvillian as
\begin{equation}
\mathcal{L}_0 = \mathcal{L}_\mathrm{s} + \mathcal{L}_\mathrm{a}
\end{equation}
with $\mathcal{L}_\mathrm{a} [\rho] = -\ii [\hat{H}_\mathrm{a}, \rho]$, 
and the Liouvillian perturbation
\begin{equation}
\mathcal{L}_V [\rho] = -\ii [\hat{V}, \rho],
\end{equation}
which is proportional to the Rabi frequency $\Omega$.

For the initial state $\rho_0 = \rho_\mathrm{s} \otimes \rho_\mathrm{a}$, 
the assumptions that $\rho_\mathrm{s}$ is stationary under $\mathcal{L}_\mathrm{s}$ and $\rho_\mathrm{a}$ is diagonal in the momentum basis
imply $\mathcal{L}_\mathrm{s} [\rho_\mathrm{s}] = 0$ and $\mathcal{L}_\mathrm{a} [\rho_\mathrm{a}] = 0$, respectively.
Therefore, the initial state $\rho_0$ is stationary under the unperturbed Liouvillian $\mathcal{L}_0$, i.e., $\mathcal{L}_0 [\rho_0] = 0$.
Since $\rho_\mathrm{a}$ is diagonal, we further have 
$\Tr_\mathrm{a} (\hat{a}_k^\dagger \hat{a}_q \rho_\mathrm{a}) = N_{\mathrm{a}, k}^0 \delta_{k q}$
and $\Tr_\mathrm{a} (\hat{O}_\mathrm{a} \rho_\mathrm{a}) = 0$ if 
$\hat{O}_\mathrm{a}$ changes the auxiliary particle number (e.g., $\hat{a}_k$ and $\hat{a}_k \hat{a}_q$).
These relations will be used in the following derivation.

We write the density matrix as $\rho (t) = \rho_0 + \delta \rho (t)$, with $\delta \rho (t)$ satisfying
\begin{equation}\label{drho_eom}
\frac{d}{d t} \delta \rho (t) = \mathcal{L}_0 [\delta \rho (t)] + \mathcal{L}_V [\rho_0] + \mathcal{L}_V [\delta \rho (t)].
\end{equation}
By expanding $\delta \rho (t) = \sum_{n \geq 1} \delta \rho^{(n)} (t)$
with $\delta \rho^{(n)} (t)$ of order $\Omega^n$,
substituting this expansion into Eq.~\eqref{drho_eom}, 
and collecting terms of the same order in $\Omega$,
we obtain the recursion relation
\begin{equation}
\delta \rho^{(n)} (t)
= \int_0^t d t' \,
e^{\mathcal{L}_0 (t - t')}
\mathcal{L}_V [\delta \rho^{(n - 1)} (t')],
\end{equation}
with $\delta \rho^{(0)} (t) = \rho_0$. 
For sufficiently small $\Omega$, we keep only the linear term and approximate $\delta \rho (t)$ by 
\begin{equation}
\delta \rho (t) = \delta \rho^{(1)} (t) + O(\Omega^2) 
\approx
\int_0^t d t' \,
e^{\mathcal{L}_0 (t - t')} \mathcal{L}_V [\rho_0].
\end{equation}

The time derivative of the auxiliary occupation can be expressed in terms of the adjoint Liouvillian as
\begin{equation}
\dot{N}_{\mathrm{a}, k} (t)
= \Tr \! \big\{
\mathcal{L}^\dagger [\hat{N}_{\mathrm{a}, k}] \rho (t)
\big\}.
\end{equation}
Since $\mathcal{L}_0^\dagger [\hat{N}_{\mathrm{a}, k}] = 0$, only the Liouvillian perturbation contributes, giving
\begin{equation}
\dot{N}_{\mathrm{a}, k} (t)
= \Tr \! \big\{
\mathcal{L}_V^\dagger [\hat{N}_{\mathrm{a}, k}] \rho (t)
\big\}
= \ii \Tr \! \big\{
[\hat{V}, \hat{N}_{\mathrm{a}, k}] \rho (t)
\big\}.
\end{equation}
Substituting Eq.~\eqref{coh_rabi_coupling}, we obtain
\begin{equation}
\dot{N}_{\mathrm{a}, k} (t)
= \frac{\ii \Omega}{2}
\Tr \! \big[
(\hat{c}_{k m}^\dagger \hat{a}_k
- \hat{a}_k^\dagger \hat{c}_{k m}) \rho (t)
\big].
\end{equation}
Using the fact that $\Tr_\mathrm{a} (\hat{a}_k \rho_\mathrm{a}) = 0$, 
we find that the contribution from $\rho_0$ vanishes, obtaining
\begin{equation}
\begin{aligned}
\dot{N}_{\mathrm{a}, k} (t)
&= \frac{\ii \Omega}{2} 
\Tr \! \big[
(\hat{c}_{k m}^\dagger \hat{a}_k
- \hat{a}_k^\dagger \hat{c}_{k m}) \delta \rho (t)
\big]
\\ & \approx
\frac{\ii \Omega}{2} 
\int_0^t d t'
\Tr \! \big\{
(\hat{c}_{k m}^\dagger \hat{a}_k
- \hat{a}_k^\dagger \hat{c}_{k m})
e^{\mathcal{L}_0 t'} \mathcal{L}_V [\rho_0]
\big\}.
\end{aligned}
\end{equation}
We then substitute $\mathcal{L}_V [\rho_0] = -\ii [\hat{V}, \rho_0]$ 
into this equation and evaluate the trace to obtain
\begin{equation}\label{dotNakt2}
\begin{aligned}
\dot{N}_{\mathrm{a}, k} (t)
\approx
\frac{\Omega^2}{2}
&\operatorname{Im}
\int_0^t d t' \, e^{\ii \omega_\mathrm{a} t'}
\\ & \times
\big[
N_{\mathrm{a}, k}^0 \chi_k^+ (t')
- (1 + \zeta N_{\mathrm{a}, k}^0) \chi_k^- (t')
\big],
\end{aligned}
\end{equation}
where we used 
$\Tr_\mathrm{a} (\hat{a}_k^\dagger \hat{a}_q \rho_\mathrm{a}) = N_{\mathrm{a}, k}^0 \delta_{k q}$, 
$\hat{a}_k \hat{a}_q^\dagger = \delta_{k q} + \zeta \hat{a}_q^\dagger \hat{a}_k$,
$\Tr_\mathrm{a} (\hat{a}_k \hat{a}_q \rho_\mathrm{a}) = 0$, 
and $[\hat{H}_\mathrm{a}, \rho_\mathrm{a}] = 0$.
For a sufficiently long time $t$, the upper limit of the integral can be replaced by infinity so that
Eq.~\eqref{dotNakt2} reduces to Eq.~\eqref{dotNakt}.
In practice, this requires $t \gtrsim 1 / \gamma_\mathrm{min}$ 
with $\gamma_\mathrm{min}$ being the smallest quasiparticle decay rate in the system.

\section{Implementation of incoherent quasiparticle spectroscopy}
\label{inc_experiment}

Here we discuss how to engineer the dissipative pulse and 
present two methods for measuring $M_{k,n}(t)$ in Eq.~\eqref{inc_measurement}.

\subsection{Engineering the dissipative pulse}

Suppose that each unit cell $j$ comprises system levels labeled by $m$ and an auxiliary level,
whose annihilation operators are denoted by $\hat{c}_{jm}$ and $\hat{a}_j$, respectively.
We introduce a short-lived excited state whose annihilation operator is given by $\hat{e}_j$.
We assume that a weak optical drive couples the system level $m$ to the excited state,
which subsequently decays to the auxiliary level via spontaneous emission.
The pump Hamiltonian and the associated jump operators describing these processes are given by
\begin{equation}
\begin{aligned}
\hat{H}_\mathrm{p} &= \frac{1}{2} \sum_j
\big(\Omega_j \hat{e}_j^\dagger \hat{c}_{jm} + \mathrm{H.c.}\big),
\\
\hat{L}_{\mathrm{e}, j} &= \sqrt{\Gamma}\,\hat{a}_j^\dagger \hat{e}_j,
\end{aligned}
\end{equation}
where $\Omega_j$ is the Rabi frequency and $\Gamma$ is the decay rate.
Specifically, by applying two laser beams whose wavevectors differ by $k$, 
one can realize a Rabi frequency of the form
\begin{equation}
\Omega_j = \Omega_\mathrm{p} (1 + e^{\ii k j}).
\end{equation}
For $|\Omega_j| \ll \Gamma$ and initially empty auxiliary levels,
adiabatic elimination of the excited state leads to a purely dissipative coupling
with jump operators~\cite{Reiter2012PRA}
\begin{equation}
\hat{L}_{\mathrm{d},j}
= \frac{\ii\Omega_j}{\sqrt{\Gamma}}
\hat{a}_j^\dagger \hat{c}_{jm}.
\end{equation}
Since the overall phase of each jump operator does not affect the dissipator,
we can equivalently write
\begin{equation}
\hat{L}_{\mathrm{d},j}
= \sqrt{\gamma_{\mathrm{d},j}}\,
\hat{a}_j^\dagger \hat{c}_{jm},
\end{equation}
where $\gamma_{\mathrm{d},j} = |\Omega_j|^2/\Gamma
= \gamma_\mathrm{d}[1+\cos(kj)]$ and $\gamma_\mathrm{d} = 2|\Omega_\mathrm{p}|^2 / \Gamma$.
This realizes the dissipative pulse generated by $\mathcal{L}_\mathrm{d}$ in Eq.~\eqref{dissipative_L_d}.

\subsection{Measurement of $M_{k,n}$}

We now discuss how to measure
$M_{k,n} = \langle \hat{N}_\mathrm{a}\hat{D}_{k,n} \rangle$,
with $\hat{N}_\mathrm{a}$ and $\hat{D}_{k,n}$ given by
Eqs.~\eqref{inc_aux_number} and \eqref{inc_density_readout}, respectively, and
$\langle \hat{O} \rangle = \Tr (\hat{O}\rho)$ being the expectation value
for an arbitrary density matrix $\rho$.
Since
\begin{equation}
M_{k,n} = \sum_{ij}\cos\!\left(kj-\frac{n\pi}{2}\right)
\big(
\langle\hat{n}_{\mathrm{a},i}\rangle
+ \zeta \langle\hat{n}_{\mathrm{a},i}\hat{n}_{jm} \rangle
\big),
\end{equation}
where $\hat{n}_{\mathrm{a},i}=\hat{a}_i^\dagger\hat{a}_i$ and
$\hat{n}_{jm}=\hat{c}_{jm}^\dagger\hat{c}_{jm}$,
a direct approach is to perform site-resolved measurements, such as
quantum gas microscopy~\cite{Bakr2009Nature,Preiss2015PRA,Fukuhara2015PRL},
to simultaneously determine the system and auxiliary occupations.
In each round of measurement $\alpha$, we record the auxiliary and system occupations
$n_{\mathrm{a},i}^{(\alpha)}$ and $n_{jm}^{(\alpha)}$ at all sites.
These values are either $0$ or $1$ for fermions, or nonnegative integers for bosons 
(obtained from projective measurements).
Then we have
\begin{equation}
M_{k,n} \approx \frac{1}{R}
\sum_{\alpha=1}^{R}
N_\mathrm{a}^{(\alpha)}
\sum_j
\cos\!\left(kj-\frac{n\pi}{2}\right)
\big(1 + \zeta n_{jm}^{(\alpha)}\big),
\end{equation}
where $R$ is the number of rounds and
$N_\mathrm{a}^{(\alpha)} = \sum_i n_{\mathrm{a},i}^{(\alpha)}$
is the total auxiliary occupation in round $\alpha$.

Alternatively, $M_{k, n}$ can be detected without relying on site-resolved measurements.
We introduce a readout level in each unit cell $j$, whose annihilation operator is given by $\hat{r}_j$.
These levels can be realized using an additional internal state of the atoms.
Starting from the initial density matrix $\rho \otimes |0\rangle_\mathrm{r} \langle 0|_\mathrm{r}$,
where $|0\rangle_\mathrm{r}$ denotes the vacuum state of all readout levels,
we apply two consecutive $\pi/2$ pulses.
These pulses couple the system level $m$ to the readout level through the Hamiltonian
\begin{equation}
\hat{H}_\mathrm{r} (\bm{\vartheta})
= \frac{\Omega_\mathrm{r}}{2} \sum_j
\big(e^{\ii \vartheta_j} \hat{r}_j^\dagger \hat{c}_{jm} + \mathrm{H.c.} \big),
\end{equation}
where $\bm{\vartheta} = (\vartheta_1, \ldots, \vartheta_L)$.
Each pulse has a duration of $t_{\pi/2} = \pi/(2 \Omega_\mathrm{r})$, 
and the phases are taken as $\vartheta_{1, j} = 0$ in the first pulse
and $\vartheta_{2, j} = k j - n \pi/2$ in the second pulse.
The evolution operator for the two pulses is given by
\begin{equation}
\label{inc_two_pulse_tev_op}
\hat{U} = e^{-\ii t_{\pi/2} \hat{H}_\mathrm{r} (\bm{\vartheta}_2)}
e^{-\ii t_{\pi/2} \hat{H}_\mathrm{r} (\bm{\vartheta}_1)}.
\end{equation}
Then we have
\begin{equation}
\label{inc_ramsey_readout}
M_{k, n}
= f_{k, n} \Tr (\hat{N}_\mathrm{a} \rho')
+ \zeta \Tr (\hat{N}_\mathrm{a} \hat{N}_\mathrm{r} \rho')
- \zeta \Tr (\hat{N}_\mathrm{a} \hat{N}_m \rho'),
\end{equation}
where $\rho' = \hat{U} (\rho \otimes |0\rangle_\mathrm{r} \langle 0|_\mathrm{r}) \hat{U}^\dagger$ 
is the density matrix after the two pulses,
$f_{k, n} = \sum_j \cos (k j - n \pi/2)$,
$\hat{N}_\mathrm{r} = \sum_j \hat{r}_j^\dagger \hat{r}_j$, and
$\hat{N}_m = \sum_j \hat{c}_{jm}^\dagger \hat{c}_{jm}$.
To measure $\hat{N}_\mathrm{a}$, $\hat{N}_\mathrm{r}$, and $\hat{N}_m$ simultaneously,
one could use state-resolved measurements.
In each round of measurement $\alpha$, we record the total particle numbers
$N_\mathrm{a}^{(\alpha)}$, $N_\mathrm{r}^{(\alpha)}$, and $N_m^{(\alpha)}$
in the auxiliary, readout, and system levels, respectively.
Averaging over $R$ rounds gives
\begin{equation}
M_{k, n} \approx \frac{1}{R} \sum_{\alpha=1}^{R}
N_\mathrm{a}^{(\alpha)} \big(f_{k, n} + \zeta N_\mathrm{r}^{(\alpha)}
- \zeta N_m^{(\alpha)} \big).
\end{equation}

We now provide a brief derivation of Eq.~\eqref{inc_ramsey_readout}.
It follows from Eq.~\eqref{inc_two_pulse_tev_op} that
\begin{equation}
\begin{aligned}
\hat{U}^\dagger (\hat{N}_\mathrm{r} - \hat{N}_m) \hat{U}
= {} & \sum_j \cos \vartheta_{2, j}
\big(\hat{n}_{jm} - \hat{r}_j^\dagger \hat{r}_j \big)
\\
& + \sum_j \sin \vartheta_{2, j}
\big(\hat{r}_j^\dagger \hat{c}_{jm} + \hat{c}_{jm}^\dagger \hat{r}_j \big).
\end{aligned}
\end{equation}
Since $\rho' = \hat{U} (\rho \otimes |0\rangle_\mathrm{r} \langle 0|_\mathrm{r}) \hat{U}^\dagger$
and $\hat{U}$ commutes with $\hat{N}_\mathrm{a}$, we have
\begin{equation}
\begin{aligned}
& \Tr \big(\hat{N}_\mathrm{a} (\hat{N}_\mathrm{r} - \hat{N}_m) \rho' \big)
\\
& = \Tr \big( \hat{N}_\mathrm{a} \hat{U}^\dagger (\hat{N}_\mathrm{r} - \hat{N}_m) \hat{U}
(\rho \otimes |0\rangle_\mathrm{r} \langle 0|_\mathrm{r}) \big)
\\
& = \sum_j \cos \vartheta_{2, j} \Tr (\hat{N}_\mathrm{a} \hat{n}_{jm} \rho),
\end{aligned}
\end{equation}
where we have used $\hat{r}_j |0\rangle_\mathrm{r} = 0$.
Together with $\Tr (\hat{N}_\mathrm{a} \rho') = \Tr (\hat{N}_\mathrm{a} \rho)$, we arrive at Eq.~\eqref{inc_ramsey_readout}:
\begin{equation}
\begin{aligned}
& \Tr \big(\hat{N}_\mathrm{a} (f_{k, n} + \zeta \hat{N}_\mathrm{r}
- \zeta \hat{N}_m) \rho' \big)
\\
& = \sum_j \cos\! \left(k j - \frac{n \pi}{2}\right)
\langle \hat{N}_\mathrm{a} (1 + \zeta \hat{n}_{jm}) \rangle
= M_{k, n}.
\end{aligned}
\end{equation}

\section{Linear response theory for incoherent quasiparticle spectroscopy}
\label{inc_derivation}

We derive the relationship between $M_{k, n} (t)$ and $X_k (t)$ in the weak pulse limit, 
as given in Eq.~\eqref{inc_lrt}.
Starting from Eq.~\eqref{incrhot1}, we obtain
\begin{equation}
\rho(t) = e^{\mathcal{L}_\mathrm{s} t} [\rho_0]
+ t_\mathrm{d} e^{\mathcal{L}_\mathrm{s} t}
\mathcal{L}_\mathrm{d}[\rho_0]
+O(\Theta_\mathrm{d}^2).
\end{equation}
Since $e^{\mathcal{L}_\mathrm{s} t}[\rho_0] = \rho_0$ contains no auxiliary particles,
we have $\hat{N}_\mathrm{a} \rho_0 = 0$; hence, this term does not contribute to $M_{k, n} (t)$.
The same holds for the anticommutator part of $\mathcal{L}_\mathrm{d}[\rho_0]$,
namely $- \sum_j \frac{1}{2}
\{ \hat{L}_{\mathrm{d}, j}^\dagger \hat{L}_{\mathrm{d}, j}, \rho_0 \}$.
Therefore, only the quantum jump term
$\sum_j \hat{L}_{\mathrm{d}, j} \rho_0 \hat{L}_{\mathrm{d}, j}^\dagger$
contributes to $M_{k, n} (t)$, yielding
\begin{equation}
\begin{aligned}
&M_{k, n} (t)
= t_\mathrm{d} \sum_j \Tr \! \big[ \hat{N}_\mathrm{a} \hat{D}_{k,n} 
e^{\mathcal{L}_\mathrm{s} t} [\hat{L}_{\mathrm{d}, j} \rho_0 \hat{L}_{\mathrm{d}, j}^\dagger] \big]
+O(\Theta_\mathrm{d}^2)
\\
&= \ii \Theta_\mathrm{d} \sum_{i j} [1 + \cos(k j)]
\cos \! \Big( k i - \frac{n\pi}{2} \Big) X_{ij} (t)
+O(\Theta_\mathrm{d}^2),
\end{aligned}
\end{equation}
with
\begin{equation}
X_{ij} (t) = -\ii \Tr \! \big[
\hat{c}_{i m} \hat{c}_{i m}^\dagger
e^{\mathcal{L}_\mathrm{s} t}
[\hat{c}_{j m} \rho_\mathrm{s} \hat{c}_{j m}^\dagger]
\big].
\end{equation}
For a translation-invariant initial state and system Liouvillian, we have 
$X_{ij} (t) = X_r (t)$ with $r = i - j$.
Then we obtain
\begin{equation}\label{Mknt}
\begin{aligned}
M_{k, n} (t) = \, & \frac{\ii \Theta_\mathrm{d} L}{2}
[1+(-1)^n \delta_{2k,0}+2\delta_{k0}]
\\ &\times
\sum_r \cos \! \Big( k r - \frac{n\pi}{2} \Big) X_r(t)
+ O(\Theta_\mathrm{d}^2),
\end{aligned}
\end{equation}
where $\delta_{2k,0}=1$ if and only if $k=0$ or $\pi$. 
Here we take $k \in [0, 2 \pi)$.

The response function satisfies
\begin{equation}
X_k(t) = \sum_r e^{-\ii k r} X_r(t).
\end{equation}
We further use the fact that $X_r(t)$ is purely imaginary, which follows from
\begin{equation}
X_{ij}^*(t) =
\ii \Tr \! \big[
e^{\mathcal{L}_\mathrm{s} t}
[\hat{c}_{j m}\rho_\mathrm{s}\hat{c}_{j m}^\dagger]
\hat{c}_{i m} \hat{c}_{i m}^\dagger
\big]
= -X_{ij}(t).
\end{equation}
Then we have
\begin{equation}
\begin{aligned}
\operatorname{Im} X_k(t) &= - \ii \sum_r \cos(k r) X_r(t),
\\
\operatorname{Re} X_k(t) &= - \ii \sum_r \sin(k r) X_r(t),
\end{aligned}
\end{equation}
which gives
\begin{equation}\label{inc_der}
\begin{aligned}
M_{k, 0}(t) &= -\frac{\Theta_\mathrm{d} L}{2}
(1 + \delta_{k\pi} + 3\delta_{k0}) \operatorname{Im} X_k(t) + O(\Theta_\mathrm{d}^2),
\\
M_{k, 1}(t) &= -\frac{\Theta_\mathrm{d} L}{2}
(1 - \delta_{k\pi} + \delta_{k0}) \operatorname{Re} X_k(t) + O(\Theta_\mathrm{d}^2).
\end{aligned}
\end{equation}
The fact that $X_r(t)$ is purely imaginary also leads to 
$\operatorname{Re} X_0 (t) = \operatorname{Re} X_\pi(t) = 0$.
Thus Eq.~\eqref{inc_der} is equivalent to Eq.~\eqref{inc_lrt}.

\section{Time evolution of the reduced density matrix}\label{appendix1}

Here, we demonstrate that for the incoherent quasiparticle spectroscopy in fermionic systems, 
the time evolution can be performed using the reduced density matrix of the system.
We define the dissipative pulse Liouvillian for the reduced density matrix as
\begin{equation}
\mathcal{L}_{\mathrm{d},\mathrm{s}} [\rho_\mathrm{s}]
= \sum_j
\Big(
\hat{\ell}_{\mathrm{d},j} \rho_\mathrm{s} \hat{\ell}_{\mathrm{d},j}^\dagger
- \frac{1}{2}
\{ \hat{\ell}_{\mathrm{d},j}^\dagger \hat{\ell}_{\mathrm{d},j}, \rho_\mathrm{s} \}
\Big),
\end{equation}
where the jump operators are
\begin{equation}
\hat{\ell}_{\mathrm{d},j}
=
\sqrt{\gamma_\mathrm{d} [1+\cos(kj)]} \,
\hat{c}_{j m}.
\end{equation}
Then we can prove that
\begin{equation}\label{eqap1}
\begin{aligned}
\Tr_\mathrm{a} [\rho(t)]
&= \Tr_\mathrm{a} \!
\big(
e^{\mathcal{L}_\mathrm{s} t}
e^{\mathcal{L}_\mathrm{d} t_\mathrm{d}}
[\rho_\mathrm{s} \otimes |0\rangle_\mathrm{a}\langle 0|_\mathrm{a}]
\big)
\\&= e^{\mathcal{L}_\mathrm{s} t}
e^{\mathcal{L}_{\mathrm{d},\mathrm{s}} t_\mathrm{d}}
[\rho_\mathrm{s}].
\end{aligned}
\end{equation}

We first note that, since $\mathcal{L}_\mathrm{s}$ acts only on the system levels, we have
\begin{equation}
\begin{aligned}
& \Tr_\mathrm{a} \!
\big(
e^{\mathcal{L}_\mathrm{s} t}
e^{\mathcal{L}_\mathrm{d} t_\mathrm{d}}
[\rho_\mathrm{s} \otimes |0\rangle_\mathrm{a}\langle 0|_\mathrm{a}]
\big)
\\
&= e^{\mathcal{L}_\mathrm{s} t} 
\big[ \!
\Tr_\mathrm{a} \!
\big(
e^{\mathcal{L}_\mathrm{d} t_\mathrm{d}}
[\rho_\mathrm{s} \otimes |0\rangle_\mathrm{a}\langle 0|_\mathrm{a}]
\big)
\big].
\end{aligned}
\end{equation}
Furthermore, by defining
\begin{equation}
\rho_\mathrm{s}'(t_\mathrm{d})
= \Tr_\mathrm{a} \!
\big(
e^{\mathcal{L}_\mathrm{d} t_\mathrm{d}}
[\rho_\mathrm{s} \otimes |0\rangle_\mathrm{a}\langle 0|_\mathrm{a}]
\big),
\end{equation}
we have
\begin{equation}
\begin{aligned}
\frac{d}{d t} \rho_\mathrm{s}'(t)
&= \Tr_\mathrm{a} \! \big(
\mathcal{L}_\mathrm{d}
e^{\mathcal{L}_\mathrm{d} t}
[\rho_\mathrm{s} \otimes |0\rangle_\mathrm{a}\langle 0|_\mathrm{a}]
\big)
\\ &=
\sum_j
\Big(
\hat{\ell}_{\mathrm{d},j} \rho_\mathrm{s}'(t) \hat{\ell}_{\mathrm{d},j}^\dagger
- \frac{1}{2}
\{ \hat{\ell}_{\mathrm{d},j}^\dagger \hat{\ell}_{\mathrm{d},j}, \rho_\mathrm{s}'(t) \}
\Big)
\\
&= \mathcal{L}_{\mathrm{d}, \mathrm{s}}
[\rho_\mathrm{s}'(t)],
\end{aligned}
\end{equation}
where we used the following relation in deriving the second equality:
\begin{equation}
\begin{aligned}
& \hat{c}_{jm} \! \Tr_\mathrm{a} \! \big( \hat{a}_j^\dagger \hat{a}_j 
e^{\mathcal{L}_\mathrm{d} t} 
[\rho_\mathrm{s} \otimes | 0 \rangle_\mathrm{a}\langle 0|_\mathrm{a}] 
\big)
\\
& = 
\Tr_\mathrm{a} \! \big( \hat{a}_j^\dagger \hat{a}_j 
e^{\mathcal{L}_\mathrm{d} t} 
[\rho_\mathrm{s} \otimes | 0 \rangle_\mathrm{a}\langle 0|_\mathrm{a}] 
\big) \hat{c}_{jm}^\dagger 
\\
& = 0.
\end{aligned}
\end{equation}
This relation can be proved by expanding $e^{\mathcal{L}_\mathrm{d} t}$ in Taylor series
(for each term, an occupied auxiliary level $j$ implies an empty system level $j, m$).
Therefore, 
\begin{equation}
\begin{aligned}
&\Tr_\mathrm{a} \! \big(
\hat{L}_{\mathrm{d}, j}
e^{\mathcal{L}_\mathrm{d} t}
[\rho_\mathrm{s} \otimes |0\rangle_\mathrm{a}\langle 0|_\mathrm{a}]
\hat{L}_{\mathrm{d}, j}^\dagger
\big)
\\&= \hat{\ell}_{\mathrm{d}, j} \! \Tr_\mathrm{a} \! \big[
\hat{a}_j \hat{a}_j^\dagger
e^{\mathcal{L}_\mathrm{d} t}
[\rho_\mathrm{s} \otimes |0\rangle_\mathrm{a}\langle 0|_\mathrm{a}]
\big] \hat{\ell}_{\mathrm{d}, j}^\dagger
\\&=
\hat{\ell}_{\mathrm{d}, j} \rho_\mathrm{s}'(t) \hat{\ell}_{\mathrm{d}, j}^\dagger
- \hat{\ell}_{\mathrm{d}, j} \! \Tr_\mathrm{a} \! \big[ \hat{a}_j^\dagger \hat{a}_j 
e^{\mathcal{L}_\mathrm{d} t}
[\rho_\mathrm{s} \otimes | 0 \rangle_\mathrm{a}\langle 0|_\mathrm{a}]
\big] \hat{\ell}_{\mathrm{d}, j}^\dagger
\\&= \hat{\ell}_{\mathrm{d}, j} \rho_\mathrm{s}'(t) \hat{\ell}_{\mathrm{d}, j}^\dagger,
\end{aligned}
\end{equation}
and similarly
\begin{equation}
\Tr_\mathrm{a} \! \big(
\{ 
\hat{L}_{\mathrm{d}, j}^\dagger \hat{L}_{\mathrm{d}, j},
e^{\mathcal{L}_\mathrm{d} t}
[\rho_\mathrm{s} \otimes |0\rangle_\mathrm{a}\langle 0|_\mathrm{a}]
\}
\big)
=
\{
\hat{\ell}_{\mathrm{d}, j}^\dagger \hat{\ell}_{\mathrm{d}, j},
\rho_\mathrm{s}'(t)
\}.
\end{equation}
Finally, using $\rho_\mathrm{s}'(0)=\rho_\mathrm{s}$, we obtain
\begin{equation}
\rho_\mathrm{s}'(t_\mathrm{d})
=
e^{\mathcal{L}_{\mathrm{d},\mathrm{s}} t_\mathrm{d}}
[\rho_\mathrm{s}],
\end{equation}
which completes the proof of Eq.~\eqref{eqap1}.

One can further prove that for a number-conserving system Liouvillian $\mathcal{L}_\mathrm{s}$ 
and an initial state $\rho_\mathrm{s}$ satisfying $\hat{N}_\mathrm{s} \rho_\mathrm{s} = N \rho_\mathrm{s}$, 
$M_{k,n}(t)$ can be calculated using the reduced density matrix as
\begin{equation}\label{eqap2}
M_{k,n}(t)
= \Tr \! \big\{ 
(N-\hat{N}_\mathrm{s}) \hat{D}_{k,n} \!
\Tr_\mathrm{a} [\rho(t)]
\big\},
\end{equation}
where $\hat{N}_\mathrm{s}$ is the total particle-number operator of the system levels.
To prove this, we use the fact that the dissipative pulse Liouvillian $\mathcal{L}_\mathrm{d}$ conserves the total particle number,
such that $\hat{N} \rho(t) = N \rho(t)$ with $\hat{N} = \hat{N}_\mathrm{s} + \hat{N}_\mathrm{a}$.
Then we have
\begin{equation}
\begin{aligned}
M_{k,n}(t)
&=
\Tr \! \big[
(\hat{N} - \hat{N}_\mathrm{s}) \hat{D}_{k,n} \rho(t)
\big]
\\
&=
\Tr \! \big[
(N-\hat{N}_\mathrm{s}) \hat{D}_{k,n} \rho(t)
\big]
\\
&=
\Tr_\mathrm{s} \! \big\{
(N-\hat{N}_\mathrm{s}) \hat{D}_{k,n} \!
\Tr_\mathrm{a}[\rho(t)]
\big\}.
\end{aligned}
\end{equation}

\section{Derivation of the effective Hamiltonian}\label{appendix2}

Here we derive the effective Hamiltonian $\tilde{H}_\mathrm{inc}^\mathrm{A} (k)$
in Eq.~\eqref{inc_modelB_TB} using a fermionic Gaussian state as a variational ansatz.
We take the density matrix to be
\begin{equation}\label{inc_gaussian_ansatz}
\rho = \frac{1}{Z}
\exp \! \Big( - \sum_{ij} K_{ij} \hat{c}_i^\dagger \hat{c}_j \Big),
\end{equation}
where $K = K^\dagger$ and $Z$ ensures $\Tr(\rho) = 1$.
We define the correlation function as
\begin{equation}\label{inc_gaussian_C}
C_{ij} = \Tr(\hat{c}_j^\dagger \hat{c}_i \rho).
\end{equation}
We employ the master equation to obtain the dynamics of $C$ as
\begin{equation}\label{inc_gaussian_C_dynamics}
\begin{aligned}
\dot{C}_{ij}
= {} & \! \Tr \! \left( \hat{c}_j^\dagger \hat{c}_i
\mathcal{L}_\mathrm{s}^\mathrm{A}[\rho] \right)
\\
= {} & \ii J_h \big( C_{i+1, j} + C_{i-1, j} - C_{i, j+1} - C_{i, j-1} \big)
\\
&- (\gamma_\mathrm{R} + \gamma_\mathrm{L}) C_{ij}
\\
&+ \delta_{ij} \big( \gamma_\mathrm{R} C_{i-1, i-1}
+ \gamma_\mathrm{L} C_{i+1, i+1} \big)
\\
&+ \frac{\gamma_\mathrm{R} - \gamma_\mathrm{L}}{2}
\big( C_{i+1, i+1} - C_{i-1, i-1}
\\
&\qquad \qquad \quad + C_{j+1, j+1} - C_{j-1, j-1} \big) C_{ij}
\\
&+ \frac{\gamma_\mathrm{R} - \gamma_\mathrm{L}}{2} \big(
C_{i, i-1} C_{i-1, j} + C_{i, j-1} C_{j-1, j}
\\
&\qquad \qquad \quad - C_{i, i+1} C_{i+1, j} - C_{i, j+1} C_{j+1, j} \big),
\end{aligned}
\end{equation}
where we have used the Wick theorem~\cite{Surace2022SciPostPhysLectNotes},
\begin{equation}\label{inc_gaussian_wick}
\Tr(\hat{c}_i^\dagger \hat{c}_j^\dagger \hat{c}_\alpha \hat{c}_\beta \rho) = C_{\beta i} C_{\alpha j} - C_{\alpha i} C_{\beta j}.
\end{equation}

For the AJ model with periodic boundary conditions, the steady-state correlations are given by
\begin{equation}\label{inc_gaussian_C_steady}
C_{ij}^\mathrm{s} = \bar{n} \delta_{ij},
\end{equation}
where $\bar{n} = N/L$ is the filling fraction.
We consider small fluctuations around the steady state and write
\begin{equation}
C_{ij} = \bar{n} \delta_{ij} + \Delta C_{ij}.
\end{equation}
Substituting this expression into Eq.~\eqref{inc_gaussian_C_dynamics}
and keeping only terms linear in $\Delta C_{ij}$, we obtain
\begin{equation}\label{inc_gaussian_linearized}
\begin{aligned}
\frac{d}{dt} \Delta C_{ij}
= {} & \ii J_h \big( \Delta C_{i+1, j} + \Delta C_{i-1, j}
\\
&\quad \ \  - \Delta C_{i, j+1} - \Delta C_{i, j-1} \big)
\\
&- (\gamma_\mathrm{R} + \gamma_\mathrm{L}) \Delta C_{ij}
\\
&+ \delta_{ij} \big[
(\gamma_\mathrm{R} - \bar{n} (\gamma_\mathrm{R} - \gamma_\mathrm{L})) \Delta C_{i-1, i-1}
\\
&\qquad \ \  + (\gamma_\mathrm{L} + \bar{n} (\gamma_\mathrm{R} - \gamma_\mathrm{L})) \Delta C_{i+1, i+1}
\big].
\end{aligned}
\end{equation}

We then write $\Delta C_{ij}$ as $\Delta C_{j, r}$ with $r = i - j$ and 
perform the Fourier transform with respect to $j$ as
\begin{equation}\label{inc_gaussian_fourier}
\Delta C_{k, r} = \frac{1}{\sqrt{L}} \sum_j e^{-\ii k j} \Delta C_{j, r}.
\end{equation}
Equation~\eqref{inc_gaussian_linearized} then becomes
\begin{equation}\label{inc_gaussian_relative_dynamics}
\begin{aligned}
\frac{d}{dt} \Delta C_{k, r}
= {} & \ii J_h \big[ (1 - e^{-\ii k}) \Delta C_{k, r+1} + (1 - e^{\ii k}) \Delta C_{k, r-1} \big]
\\
&- (\gamma_\mathrm{R} + \gamma_\mathrm{L}) \Delta C_{k, r}
\\
&+ \delta_{r0} \big[
(\gamma_\mathrm{R} - \bar{n} (\gamma_\mathrm{R} - \gamma_\mathrm{L})) e^{-\ii k}
\\
&\qquad \ \  + (\gamma_\mathrm{L} + \bar{n} (\gamma_\mathrm{R} - \gamma_\mathrm{L})) e^{\ii k}
\big] \Delta C_{k, 0}.
\end{aligned}
\end{equation}
We define
\begin{equation}\label{inc_gaussian_vector}
\Delta \boldsymbol{C}_k
= (\Delta C_{k, 0}, \Delta C_{k, 1}, \ldots, \Delta C_{k, L-1})^T.
\end{equation}
Then we have
\begin{equation}\label{inc_gaussian_full_dynamics}
\ii \frac{d}{dt} \Delta \boldsymbol{C}_k
= H_\mathrm{inc}^\mathrm{A} (k) \Delta \boldsymbol{C}_k,
\end{equation}
where
\begin{equation}\label{inc_gaussian_full_H}
H_\mathrm{inc}^\mathrm{A} (k)
= \begin{pmatrix}
\ii \tilde{h}(k) & t_k & 0 & \cdots & t_k^*
\\
t_k^* & -\ii \Gamma & t_k & \ddots & 0
\\
0 & t_k^* & \ddots & \ddots & \vdots
\\
\vdots & \ddots & \ddots & \ddots & t_k
\\
t_k & 0 & \cdots & t_k^* & -\ii \Gamma
\end{pmatrix},
\end{equation}
with $\Gamma = \gamma_\mathrm{L} + \gamma_\mathrm{R}$, 
$t_k = J_h (e^{-\ii k} - 1)$, and
\begin{equation}\label{inc_gaussian_h}
\begin{aligned}
\tilde{h}(k)
= {} & \big[ \gamma_\mathrm{L} + \bar{n} (\gamma_\mathrm{R} - \gamma_\mathrm{L}) \big] (e^{\ii k} - 1)
\\
&+ \big[ \gamma_\mathrm{R} - \bar{n} (\gamma_\mathrm{R} - \gamma_\mathrm{L}) \big] (e^{-\ii k} - 1).
\end{aligned}
\end{equation}

The Hamiltonian $H_\mathrm{inc}^\mathrm{A} (k)$ governs the evolution of the perturbation within the Gaussian ansatz.
Specifically, suppose that the perturbed density matrix is given by
\begin{equation}
\rho = \frac{\rho_\mathrm{s} + \epsilon \, \mathcal{C}_{-k}[\rho_\mathrm{s}]}{1 + \epsilon \sqrt{L} \bar{n} \delta_{k0}},
\end{equation}
where $\rho_\mathrm{s}$ is the Gaussian steady state characterized by Eq.~\eqref{inc_gaussian_C_steady}, 
$\mathcal{C}_{-k} [\rho_\mathrm{s}]$ is the perturbation,
and the denominator guarantees that $\Tr (\rho) = 1$ for all $k$.
Then we have
\begin{equation}\label{inc_gaussian_initial_correlations}
\begin{aligned}
\Delta C_{q, r}
&= \frac{1}{\sqrt{L}} \sum_j e^{-\ii q j}
\Tr \! \big[ \hat{c}_j^\dagger \hat{c}_{j+r} (\rho - \rho_\mathrm{s}) \big]
\\&= \big[ - \epsilon \bar{n}^2  + O(\epsilon^2) \big] \delta_{qk} \delta_{r0}.
\end{aligned}
\end{equation}
Therefore, only $\Delta C_{k, 0}$ is initially nonzero.
To further simplify the dynamics while retaining its main features,
we consider only $\Delta C_{k, 0}$ and its nearest neighbors
$\Delta C_{k, \pm 1}$ and neglect the correlations at larger $|r|$.
The dynamics in the truncated space is then given by
\begin{equation}\label{inc_gaussian_truncated}
\ii \frac{d}{dt}
\begin{pmatrix}
\Delta C_{k, 0}
\\
\Delta C_{k, 1}
\\
\Delta C_{k, -1}
\end{pmatrix}
= \begin{pmatrix}
\ii \tilde{h}(k) & t_k & t_k^*
\\
t_k^* & -\ii \Gamma & 0
\\
t_k & 0 & -\ii \Gamma
\end{pmatrix}
\begin{pmatrix}
\Delta C_{k, 0}
\\
\Delta C_{k, 1}
\\
\Delta C_{k, -1}
\end{pmatrix}.
\end{equation}
One may further define
\begin{equation}\label{inc_gaussian_components}
\begin{aligned}
u_k &= \frac{1}{\sqrt{2}} \big( \Delta C_{k, 1} - e^{\ii k} \Delta C_{k, -1} \big),
\\
v_k &= \frac{1}{\sqrt{2}} \big( \Delta C_{k, 1} + e^{\ii k} \Delta C_{k, -1} \big).
\end{aligned}
\end{equation}
Then we have
\begin{equation}\label{inc_gaussian_transformed_dynamics}
\ii \frac{d}{dt}
\begin{pmatrix}
\Delta C_{k, 0}
\\
u_k
\\
v_k
\end{pmatrix}
= \begin{pmatrix}
\ii \tilde{h}(k) & \sqrt{2} t_k & 0
\\
\sqrt{2} t_k^* & -\ii \Gamma & 0
\\
0 & 0 & -\ii \Gamma
\end{pmatrix}
\begin{pmatrix}
\Delta C_{k, 0}
\\
u_k
\\
v_k
\end{pmatrix}.
\end{equation}
Here $u_k$ couples to $\Delta C_{k, 0}$ and $v_k$ decays independently at rate $\Gamma$.
Since $v_k$ is initially zero, it remains zero during the time evolution.
The dynamics of $(\Delta C_{k, 0}, u_k)^T$ is therefore governed by
\begin{equation}
\tilde{H}_\mathrm{inc}^\mathrm{A} (k)
= \begin{pmatrix}
\ii \tilde{h}(k) & \sqrt{2} t_k
\\
\sqrt{2} t_k^* & -\ii \Gamma
\end{pmatrix},
\end{equation}
which gives Eq.~\eqref{inc_modelB_TB}.

\section{Mean-field analysis of the weak $U(1)$ symmetric model}
\label{appendix_weak_u1_mf}

Here we derive the mean-field critical point and steady-state order parameters given in
Eqs.~\eqref{weak_u1_mf_boundary} and~\eqref{weak_u1_mf_order}
for the model in Sec.~\ref{subsec_weak_u1}.
We use the product state
\begin{equation}
\rho_\mathrm{MF}
= \bigotimes_{i=1}^L\left(\rho_{iA}\otimes\rho_{iB}\right)
\end{equation}
as the mean-field ansatz, with $\rho_{i\nu}$ being the local density matrix at site $(i,\nu)$.
We define the local fields and densities by
\begin{equation}
\label{wu1_local_fields}
\begin{aligned}
\psi_{i\nu}
&= \langle \hat{b}_{i\nu} \rangle
= \Tr(\rho_{i\nu} \hat{b}_{i\nu}),
\\
n_{i\nu}
&= \langle \hat{n}_{i\nu} \rangle
= \Tr(\rho_{i\nu} \hat{n}_{i\nu}),
\end{aligned}
\end{equation}
where $\hat{n}_{i\nu} = \hat{b}_{i\nu}^\dagger \hat{b}_{i\nu}$ and $\nu=A, B$.
Within the mean-field approximation, 
we obtain the differential equations derived from the master equation as
\begin{equation}
\label{wu1_eom}
\begin{aligned}
\dot{\psi}_{iA}
&= -\frac{\gamma}{2} \psi_{iA}
- \ii (1 - 2 n_{iA}) \sum_j h_{ij} \psi_{jB},
\\
\dot{\psi}_{iB}
&= -\frac{\gamma}{2} \psi_{iB}
- \ii (1 - 2 n_{iB}) \sum_j h_{ji} \psi_{jA},
\\
\dot{n}_{iA}
&= \gamma(1-n_{iA})
+ 2 \sum_j h_{ij} \operatorname{Im} (\psi_{iA}^* \psi_{jB}),
\\
\dot{n}_{iB}
&= -\gamma n_{iB}
- 2 \sum_j h_{ji} \operatorname{Im} (\psi_{jA}^* \psi_{iB}),
\end{aligned}
\end{equation}
where
\begin{equation}
\label{wu1_h}
h_{ij} = \frac{J}{|2(j-i)+1|^{p}}.
\end{equation}

To find the mean-field steady states, we further utilize the translation symmetry
such that $\psi_{i\nu} = \psi_{\nu}$ and $n_{i\nu} = n_{\nu}$.
The steady states are then the solutions of the following equations:
\begin{equation}
\begin{aligned}
0 &= - \frac{\gamma}{2} \psi_A
- \ii K_{p} (1 - 2 n_A) \psi_B,
\\
0 &= - \frac{\gamma}{2} \psi_B
- \ii K_{p} (1 - 2 n_B) \psi_A,
\\
0 &= \gamma (1 - n_A)
+ 2 K_{p} \operatorname{Im} (\psi_A^* \psi_B),
\\
0 &= - \gamma n_B
- 2 K_{p} \operatorname{Im} (\psi_A^* \psi_B),
\end{aligned}
\end{equation}
where
\begin{equation}
\label{wu1_K}
K_{p}
= \sum_j h_{ij} = \sum_j h_{ji}
= 2 J (1 - 2^{-p}) \zeta (p),
\end{equation}
with $\zeta (p)$ being the Riemann zeta function.

We consider $J > 0$ and $\gamma > 0$, and obtain two types of mean-field steady states.
One is the weak $U(1)$ symmetric solution given by
\begin{equation}
\psi_A = \psi_B = 0,
\qquad
n_A = 1,
\qquad
n_B = 0,
\end{equation}
which exists for all $J > 0$.
The other is the symmetry-broken solution given by
\begin{equation}
\label{wu1_ss}
\begin{aligned}
& \psi_A = \ii \psi_B = e^{\ii \phi} \sqrt{\frac{\eta (1 - \eta)}{2}},
\\
& n_A = \frac{1 + \eta}{2},
\qquad
n_B = \frac{1 - \eta}{2},
\end{aligned}
\end{equation}
with $\eta = \gamma / (2 K_{p})$
and $\phi$ being an arbitrary phase.
This solution exists for $J > \gamma / [4 (1 - 2^{-p}) \zeta (p)]$
and spontaneously breaks the weak $U(1)$ symmetry due to nonzero $\psi_A$ and $\psi_B$.

We further note that for $J > \gamma / [4 (1 - 2^{-p}) \zeta (p)]$, 
the symmetric solution is unstable.
By considering small perturbations to the fields around the symmetric solution 
and neglecting terms beyond linear order, we obtain
\begin{equation}
\frac{d}{dt}
\begin{pmatrix}
\delta \psi_A \\
\delta \psi_B
\end{pmatrix}
=
\begin{pmatrix}
- \gamma / 2 & \ii K_{p} \\
- \ii K_{p} & - \gamma / 2
\end{pmatrix}
\begin{pmatrix}
\delta \psi_A \\
\delta \psi_B
\end{pmatrix}.
\end{equation}
For $J > \gamma / [4 (1 - 2^{-p}) \zeta (p)]$, 
a generic small perturbation grows exponentially 
and drives the system away from the symmetric solution.

\subsection{Dispersion relation}
\label{wu1_fluct}

Here we derive the mean-field dispersion relation of the Nambu--Goldstone mode
associated with the spontaneous breaking of the weak $U(1)$ symmetry.
In the symmetry-broken phase, we write the perturbed local fields as
\begin{equation}
\label{wu1_phase}
\psi_{iA} = \psi_{\mathrm{abs}} e^{\ii \theta_{iA}},
\qquad
\psi_{iB} = - \ii \psi_{\mathrm{abs}} e^{\ii \theta_{iB}},
\end{equation}
where $\psi_{\mathrm{abs}} = \sqrt{\eta (1 - \eta) / 2}$,
$\theta_{i\nu}$ is the local phase perturbation,
and we choose the global phase to be $\phi = 0$.
Substituting these into Eq.~\eqref{wu1_eom}
with the local densities fixed at their steady-state values in Eq.~\eqref{wu1_ss} 
and keeping only terms linear in $\theta$,
we obtain
\begin{equation}
\label{wu1_phase_eom}
\begin{aligned}
\dot{\theta}_{iA}
&= - \frac{\gamma}{2} \theta_{iA}
+ \eta \sum_j h_{ij} \theta_{jB},
\\
\dot{\theta}_{iB}
&= - \frac{\gamma}{2} \theta_{iB}
+ \eta \sum_j h_{ji} \theta_{jA}.
\end{aligned}
\end{equation}
We substitute the Fourier transform
\begin{equation}
\theta_{k\nu} = \frac{1}{\sqrt{L}} \sum_j e^{- \ii k j} \theta_{j\nu}
\end{equation}
in Eq.~\eqref{wu1_phase_eom} to obtain
\begin{equation}
\label{wu1_mat}
\frac{d}{dt}
\begin{pmatrix}
\theta_{kA} \\
\theta_{kB}
\end{pmatrix}
=
\begin{pmatrix}
- \gamma / 2 & \eta h_k \\
\eta h_k^* & - \gamma / 2
\end{pmatrix}
\begin{pmatrix}
\theta_{kA} \\
\theta_{kB}
\end{pmatrix},
\end{equation}
where
\begin{equation}
h_k
= J \sum_{r = -\infty}^{\infty}
\frac{e^{\ii k r}}{|2 r + 1|^{p}}.
\end{equation}
The eigenvalues of the matrix in Eq.~\eqref{wu1_mat} are
\begin{equation}
\label{wu1_eig}
\lambda_\pm (k) = - \frac{\gamma}{2} \pm \eta |h_k|.
\end{equation}
Since $h_0 = K_{p}$ and $\eta K_{p} = \gamma / 2$,
we have $\lambda_+ (0) = 0$ and $\lambda_- (0) = - \gamma$.
Thus the $\lambda_+$ branch describes the Nambu--Goldstone mode associated with the spontaneous breaking
of the weak $U(1)$ symmetry.

To obtain the long-wavelength dispersion, we use Eq.~(20) in
Ref.~\cite{Bruno2001PRL}, which in one dimension reads
\begin{equation}
\begin{aligned}
\tilde{E}_{p}(q)
= \sum_{n \neq 0} \frac{1 - \cos(n q)}{|n|^{p}}
\propto
\begin{cases}
|q|^{p - 1}, & 1 < p < 3, \\
q^2 \log (1 / |q|), & p = 3, \\
q^2, & p > 3
\end{cases}
\end{aligned}
\end{equation}
as $q \to 0$.
We further have
\begin{equation}
\begin{aligned}
h_k
&= 2 J e^{- \ii k / 2} \sum_{r = 0}^{\infty}
\frac{\cos [(r + 1/2) k]}{(2 r + 1)^{p}}
\\
&= e^{- \ii k / 2} 
\big\{ h_0 - J
[ \tilde{E}_{p} (k / 2) - 2^{-p} \tilde{E}_{p} (k) ] \big\}.
\end{aligned}
\end{equation}
For sufficiently small $|k|$, we have $|h_k| = e^{\ii k / 2} h_k$.
Using $\eta h_0 = \gamma / 2$, we obtain
\begin{equation}
\label{wu1_disp}
\begin{aligned}
-\lambda_+ (k)
&= J \eta [ \tilde{E}_{p} (k / 2) - 2^{-p} \tilde{E}_{p} (k) ]
\\
&\propto
\begin{cases}
|k|^{p - 1}, & 1 < p < 3, \\
k^2 \log (1 / |k|), & p = 3, \\
k^2, & p > 3
\end{cases}
\end{aligned}
\end{equation}
for $k \to 0$.

\subsection{Long-range order}
\label{wu1_cluster}

In one- and two-dimensional systems with short-range interactions,
fluctuations of the order parameter can destroy $U(1)$ long-range order
at nonzero temperatures~\cite{Mermin1966PRL, Hohenberg1967PR}.
However, long-range interactions can support $U(1)$ long-range order
at nonzero temperatures in certain classical systems~\cite{Kosterlitz1976PRL, Frohlich1978CMP}.
Here we employ cluster mean-field and many-body calculations to examine whether
the long-range coupling in our model allows long-range order to survive.

We divide the chain into clusters $\mathcal{C}$, each containing
$N_{\mathrm{uc}}$ consecutive unit cells, and use the product state
\begin{equation}
\label{wu1_cluster_ansatz}
\rho_\mathrm{CMF} = \bigotimes_{\mathcal{C}} \bar{\rho}_{\mathcal{C}}
\end{equation}
as the cluster mean-field ansatz for numerical calculations~\cite{Jin2016PRX}.
We assume translation invariance so that all clusters have the same density matrix $\bar{\rho}_{\mathcal{C}}$.
Starting from the master equation, we perform a partial trace over
all sites outside the cluster $\mathcal{C}$ and obtain
\begin{equation}
\label{wu1_cluster_eom}
\dot{\bar{\rho}}_{\mathcal{C}}
= -\ii [\hat{H}_{\mathcal{C}}^\mathrm{MF}, \bar{\rho}_{\mathcal{C}}]
+ \gamma \sum_{i \in \mathcal{C}}
\big( \mathcal{D}[\hat{b}_{iA}^\dagger]
+ \mathcal{D}[\hat{b}_{iB}] \big) \bar{\rho}_{\mathcal{C}},
\end{equation}
where
\begin{equation}
\label{wu1_dissipator}
\mathcal{D}[\hat{o}] \rho
= \hat{o} \rho \hat{o}^\dagger
- \frac{1}{2} \{ \hat{o}^\dagger \hat{o}, \rho \},
\end{equation}
and the mean-field Hamiltonian is
\begin{equation}
\label{wu1_cluster_h}
\hat{H}_{\mathcal{C}}^\mathrm{MF}
= \sum_{i,j \in \mathcal{C}} h_{ij}
\hat{b}_{iA}^\dagger \hat{b}_{jB}
+ \sum_{i \in \mathcal{C},\nu}
F_{i\nu} \hat{b}_{i\nu}^\dagger + \mathrm{H.c.},
\end{equation}
with
\begin{equation}
\label{wu1_cluster_fields}
F_{iA} = \sum_{j \notin \mathcal{C}} h_{ij} \psi_{jB},
\qquad
F_{iB} = \sum_{j \notin \mathcal{C}} h_{ji} \psi_{jA}.
\end{equation}

\begin{figure}[t]
\centering
\includegraphics[width=\columnwidth]{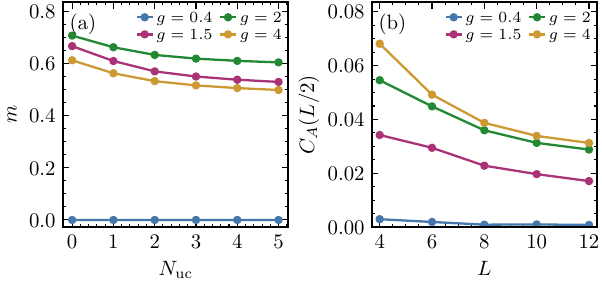}
\caption{(a) Order parameter $m$ versus the number of unit cells $N_{\mathrm{uc}}$ per cluster.
(b) Many-body steady-state correlation function $C_A(L/2)$ versus system size $L$ under periodic boundary conditions,
averaged over 4096 steady-state samples from 1024 trajectories.
Different lines correspond to different values of $g = J / J_\mathrm{c}^\mathrm{MF}$.
In (a), the points at $N_{\mathrm{uc}} = 0$ represent the results obtained from the single-site mean-field theory [Eq.~\eqref{wu1_ss}].
Here $\gamma = 1$ and $p = 5/4$.
}
\label{figS1}
\end{figure}

We choose the initial state to be
\begin{equation}
\label{wu1_cluster_initial}
\bar{\rho}_{\mathcal{C}} (0)
= |\varphi_{\mathcal{C},0}\rangle\langle\varphi_{\mathcal{C},0}|,
\end{equation}
where
\begin{equation}
|\varphi_{\mathcal{C},0}\rangle
= \bigotimes_{i \in \mathcal{C},\nu}
\frac{|0\rangle_{i\nu} + |1\rangle_{i\nu}}{\sqrt{2}}.
\end{equation}
Then we evolve it according to Eq.~\eqref{wu1_cluster_eom}
and calculate the order parameter
\begin{equation}
\label{wu1_cluster_order}
m = \frac{1}{N_{\mathrm{uc}}} \sum_{i \in \mathcal{C},\nu}
\big| \! \Tr (\bar{\rho}_{\mathcal{C}} \hat{b}_{i\nu}) \big|.
\end{equation}

Our numerical calculation shows that, for $p = 5/4$,
the order parameter remains nonzero in a regime with $J \gtrsim J_\mathrm{c}^\mathrm{MF}$
as the cluster size increases [see Fig.~\ref{figS1}(a)].

We further examine long-range order by directly calculating the steady state of 
the many-body Liouvillian $\mathcal{L}_\mathrm{s}^\mathrm{W}$ in Eq.~\eqref{weak_u1_liouvillian}.
Specifically, we calculate the many-body correlation function
\begin{equation}
\label{wu1_many_body_correlation}
C_{\nu}(r) = \frac{1}{L} \sum_{j=1}^L
\Tr \! \big( \rho_\mathrm{s} \hat{b}_{j\nu}^\dagger \hat{b}_{j+r,\nu} \big),
\end{equation}
where $\rho_\mathrm{s}$ is the steady state of $\mathcal{L}_\mathrm{s}^\mathrm{W}$.
Using quantum jump evolution, we calculate $C_A(L/2)$ up to $L = 12$ [see Fig.~\ref{figS1}(b)].
For $g=1.5, 2, 4$, the correlation function remains large,
while for $g=0.4$, the correlation function is close to zero.
These results are consistent with the cluster mean-field analysis,
suggesting that sufficiently long-range coupling (e.g., $p = 5/4$) can support long-range order in this system.

\section{Mean-field analysis of the strong $U(1)$ symmetric model}
\label{su1_mf}

\begin{figure}[!t]
\centering
\includegraphics[width=\columnwidth]{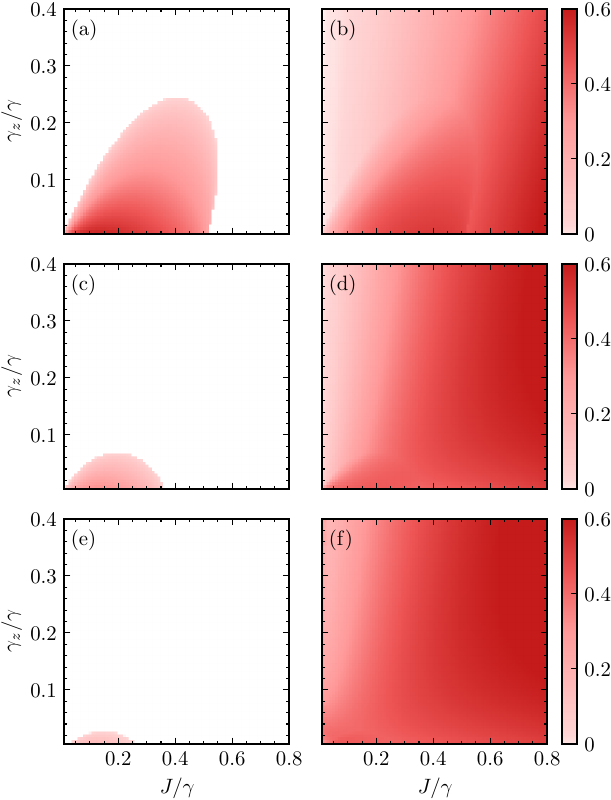}
\caption{Cluster mean-field phase diagrams for $p = 3/2$.
(a,c,e) The ordinary order parameter $m$ and
(b,d,f) the R\'enyi-2 order parameter $m^{(2)}$
as a function of $J / \gamma$ and $\gamma_z / \gamma$.
The top, middle, and bottom rows correspond to (a,b) $N_{\mathrm{uc}} = 1$,
(c,d) $N_{\mathrm{uc}} = 2$, and (e,f) $N_{\mathrm{uc}} = 3$.
}
\label{figS2}
\end{figure}

Here we perform a mean-field analysis of the model in Sec.~\ref{subsec_strong_u1}.
We follow the same procedure as in Appendix~\ref{wu1_cluster}
and obtain the mean-field dynamics of the cluster density matrix as
\begin{equation}
\label{su1_cluster_eom}
\begin{aligned}
&\dot{\bar{\rho}}_{\mathcal{C}}
= -\ii [\hat{H}_{\mathcal{C}}^\mathrm{MF}, \bar{\rho}_{\mathcal{C}}]
+ \gamma_z \sum_{i \in \mathcal{C},\nu}
\mathcal{D}[\hat{\sigma}_{i\nu}^z] \bar{\rho}_{\mathcal{C}}
\\
&\quad+ \gamma \Big[
\sum_{i=1}^{N_{\mathrm{uc}}} \mathcal{D}[\hat{b}_{iB}^\dagger \hat{b}_{iA}]
+ \sum_{i=2}^{N_{\mathrm{uc}}} \mathcal{D}[\hat{b}_{i-1,B}^\dagger \hat{b}_{iA}]
\Big] \bar{\rho}_{\mathcal{C}}
\\
&\quad+ \gamma \Big[
(1 - n_{N_{\mathrm{uc}},B}) \mathcal{D}[\hat{b}_{1A}]
+ n_{1A} \mathcal{D}[\hat{b}_{N_{\mathrm{uc}},B}^\dagger]
\Big] \bar{\rho}_{\mathcal{C}},
\end{aligned}
\end{equation}
where we assume half filling,
$N_{\mathrm{uc}}^{-1} \sum_{i \in \mathcal{C}} (n_{iA} + n_{iB}) = 1$.
Here the unit cells in $\mathcal{C}$ are labeled by $i = 1, \ldots, N_{\mathrm{uc}}$,
and $\hat{H}_{\mathcal{C}}^\mathrm{MF}$ is given in Eq.~\eqref{wu1_cluster_h}.

We use the initial state in Eq.~\eqref{wu1_cluster_initial}
and evolve it according to Eq.~\eqref{su1_cluster_eom}.
Then we calculate the ordinary order parameter $m$ 
defined in Eq.~\eqref{wu1_cluster_order}
and the R\'enyi-2 order parameter
\begin{equation}
m^{(2)} = \frac{1}{N_{\mathrm{uc}}} \sum_{i \in \mathcal{C},\nu}
\frac{\Tr (\bar{\rho}_{\mathcal{C}} \hat{b}_{i\nu}
\bar{\rho}_{\mathcal{C}} \hat{b}_{i\nu}^\dagger)}
{\Tr (\bar{\rho}_{\mathcal{C}}^2)}.
\end{equation}
For $m^{(2)} \neq 0$ and $m = 0$, the system exhibits strong-to-weak $U(1)$ SSB.
The remaining weak $U(1)$ symmetry is spontaneously broken when $m$ also becomes nonzero.

Figure~\ref{figS2} shows the results for $p = 3/2$.
For all parameter values and cluster sizes considered, we find nonzero $m^{(2)}$ [see Fig.~\ref{figS2}(b,d,f)], 
showing that the strong $U(1)$ symmetry is either broken into a weak one or fully broken.
For $N_{\mathrm{uc}} = 1$, we also find a broad regime with nonzero $m$ [see Fig.~\ref{figS2}(a)].
However, this regime shrinks significantly as $N_{\mathrm{uc}}$ increases [see Fig.~\ref{figS2}(c,e)], 
and whether there exists a regime with additional weak $U(1)$ SSB remains unclear.

\section{Incoherent response at $k = 0$}
\label{inc_k0}

\begin{figure}[!t]
\centering
\includegraphics[width=\columnwidth]{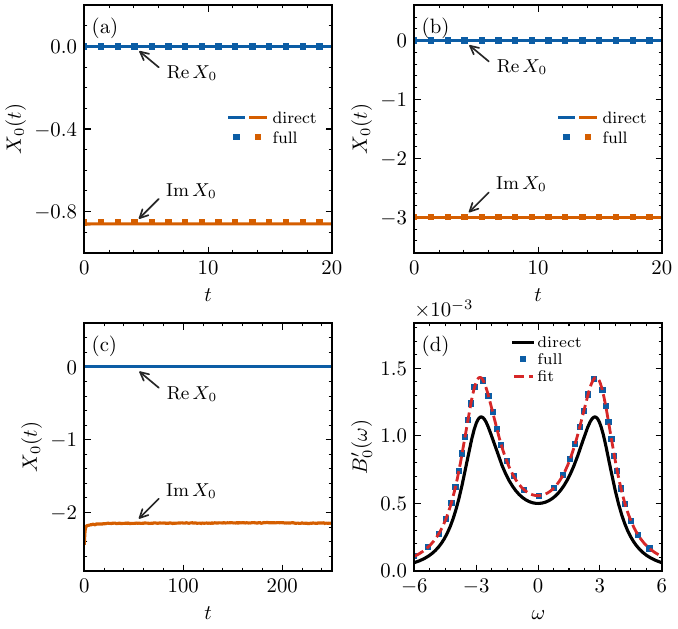}
\caption{(a)--(c) Incoherent response $X_0(t)$ for (a) the LRM model,
(b) the AJ model, and (c) the hard-core boson model discussed in Sec.~\ref{subsec_strong_u1}.
Blue and orange colors indicate the real and imaginary parts, respectively.
The solid lines represent direct calculations,
while the squares in (a) and (b) represent full simulations.
The parameters in (a) and (b) are identical to those in Figs.~\ref{fig2}(c) and~\ref{fig2}(d), respectively.
For (c), we use the parameters from Fig.~\ref{fig4}(a) with $L=10$.
(d) Spectral function $B_0'(\omega)$ for the LRM model,
where the black line and blue squares are obtained from a direct calculation
and a full simulation, respectively,
and the red dashed line is a two-peak fit to the full simulation results using Eq.~\eqref{fiteq}.
}
\label{figS3}
\end{figure}

Here we discuss the incoherent response at $k=0$ for the LRM model, the AJ model,
and the hard-core boson model introduced in Sec.~\ref{subsec_strong_u1}.
Since all these models have strong $U(1)$ symmetry,
one can consider the strong $U(1)$ symmetric sectors $\mathcal{H}_N \otimes \mathcal{H}_N$ 
in the doubled Hilbert space. Here, $\mathcal{H}_N = \operatorname{span}\{ |\psi\rangle \mid
\hat{N}_{\mathrm{s}}|\psi\rangle = N|\psi\rangle \}$
is the Hilbert space with $N$ particles.
Each of these sectors contains at least one steady state.
Assuming that $\rho_\mathrm{s}$ lies in the $N$-particle sector,
the perturbation $\mathcal{C}_{0m} [\rho_\mathrm{s}]$ belongs to the $(N-1)$-particle sector and can be expressed as
\begin{equation}\label{inc_k0_decomposition}
\mathcal{C}_{0m} [\rho_\mathrm{s}]
= \Tr\!\big(\mathcal{C}_{0m} [\rho_\mathrm{s}]\big)\rho_\mathrm{s}' + \delta\rho,
\end{equation}
where $\rho_\mathrm{s}'$ is a steady state in the $(N-1)$-particle sector satisfying $\Tr (\rho_\mathrm{s}') = 1$
and $e^{\mathcal{L}_\mathrm{s}t} [\delta \rho] \to 0$ as $t\to\infty$.
The long-time response is then given by
\begin{equation}\label{inc_k0_long_time}
X_0(t\to\infty)
= -\ii \Tr \! \big(\mathcal{C}_{0m} [\rho_\mathrm{s}]\big)
\Tr \! \big(\mathcal{C}_{0m}^\dagger [\rho_\mathrm{s}']\big),
\end{equation}
which is generally nonzero.
Indeed, all three models exhibit nonzero $X_0(t\to\infty)$ [see Fig.~\ref{figS3}(a--c)],
in contrast to the response at $k \neq 0$ which decays to 0 [see Fig.~\ref{fig2}(a)].
This nonzero long-time value gives a $\delta(\omega)$ contribution to $B_0(\omega)$,
corresponding to an incoherent quasiparticle with energy $z=0$.

When fitting the incoherent quasiparticle spectra, 
we treat the $k = 0$ point separately from other $k$ points.
Specifically, for the LRM model, we fix one quasiparticle energy to $z=0$ and fit
\begin{equation}\label{inc_k0_lrm_residual}
B_0'(\omega)
= -2\operatorname{Im}\int_0^\infty dt\,e^{\ii\omega t}
\big[X_0(t)-X_0(\infty)\big]
\end{equation}
using Eq.~\eqref{fiteq} to obtain the other two quasiparticle energies shown in Fig.~\ref{fig2}(c) [see Fig.~\ref{figS3}(d)].
For the AJ model and the hard-core boson model introduced in Sec.~\ref{subsec_strong_u1},
we assume there is only one quasiparticle energy at $k=0$ and fix it to $z=0$.
For the AJ model, there are two branches at $k \neq 0$ but only one energy at $k=0$.
To form the loop in Fig.~\ref{fig2}(d), we attribute the $k=0$ point to the first branch, 
connect the first branch at $k=2\pi(L-1)/L$ to the second branch at $k = 2\pi/L$, 
and the second branch at $k=2\pi(L-1)/L$ to the first branch at $k=0$.

\end{document}